\documentclass[twocolumn,superscriptaddress,nobalancelastpage,10pt,pra,letter]{revtex4}

\usepackage{amsmath}
\usepackage{amsthm}
\usepackage{amssymb}
\usepackage{amsfonts}
\usepackage{graphicx}
\usepackage{wasysym}
\usepackage{mathrsfs}
\usepackage{yfonts}
\usepackage{bbold}
\usepackage{verbatim}
\usepackage{subfigure}
\usepackage{color}
\usepackage{caption}
\usepackage{soul} 
\usepackage{lipsum}

\newcommand{\ket}[1]{| #1 \rangle}
\newcommand{\bra}[1]{\langle #1 |}

\newcommand{\avgr}[2]{\text{tr}\left\{ \hat{\rho}_{#1} #2 \right\}}
\newcommand{\dm}{\hat{\rho}}
\newcommand{\kbra}[2]{\ket{#1}\bra{#2}}
\newcommand{\opa}{\hat{a}}

\newcommand{\opEps}{\hat{E}^{(+)}}
\newcommand{\opEns}{\hat{E}^{(-)}}

\newcommand{\I}{\mathscr{I}}

\newcommand{\sy}{\hat{\sigma}_y}
\renewcommand{\a}{\alpha}
\renewcommand{\b}{\beta}

\makeatletter
\newcommand*{\rom}[1]{\expandafter\@slowromancap\romannumeral #1@}
\makeatother

\begin{document}


\title{Characterizing mixed state entanglement through single-photon interference}

\author{Mayukh Lahiri}
\email{mlahiri@okstate.edu} \affiliation{Department of Physics,
Oklahoma State University, Stillwater, Oklahoma, USA}
\author{Radek Lapkiewicz}
\affiliation{Institute of Experimental Physics, Faculty of Physics,
University of Warsaw, Pasteura 5, Warsaw 02-093, Poland}
\author{Armin Hochrainer}
\affiliation{Vienna Center for Quantum Science and Technology (VCQ),
Faculty of Physics, University of Vienna, Boltzmanngasse 5, Vienna
A-1090, Austria.} \affiliation{Institute for Quantum Optics and
Quantum Information, Austrian Academy of Sciences, Boltzmanngasse 3,
Vienna A-1090, Austria.}
\author{Gabriela Barreto Lemos}
\email{gabibl@if.ufrj.br} 
\affiliation{Physics Department, University of Massachusetts Boston,100 Morrissey Boulevard, Boston MA 02125, USA.}\affiliation{Institute for Quantum Optics and Quantum Information,
Austrian Academy of Sciences, Boltzmanngasse 3, Vienna A-1090, Austria.}
\author{Anton Zeilinger}
\email{anton.zeilinger@univie.ac.at} \affiliation{Vienna Center for
Quantum Science and Technology (VCQ), Faculty of Physics, University
of Vienna, Boltzmanngasse 5, Vienna A-1090, Austria.}
\affiliation{Institute for Quantum Optics and Quantum Information,
Austrian Academy of Sciences, Boltzmanngasse 3, Vienna A-1090,
Austria.}

\begin{abstract}
Entanglement verification and measurement is essential for
experimental tests of quantum mechanics and also for quantum
communication and information science. Standard methods of verifying
entanglement in a bipartite mixed state require detection of both
particles and involve coincidence measurement. We present a method that enables us to verify and measure
entanglement in a two-photon mixed state without detecting one of the photons, i.e., without performing any coincidence measurement or postselection. We consider two identical sources, each of which can
generate the same two-photon mixed state but they never emit
simultaneously. We show that one can produce a set of single-photon interference
patterns, which contain information about entanglement in the
two-photon mixed state. We
prove that it is possible to retrieve the information about
entanglement from the visibility of the interference patterns. Our method reveals a distinct avenue for verifying and measuring entanglement in mixed states.

\end{abstract}

\maketitle

\section{Introduction}\label{Sec:intro}

Entanglement is a fascinating trait of quantum mechanics\textemdash in addition to its implications for the foundations of quantum mechanics, today entanglement is a key resource in quantum information science. Verification and measurement of entanglement in a quantum state is an ever growing field of research \cite{GT-ent-det-Phrep-2009,horodecki2009review}. Entanglement in two-particle (more generally bipartite) quantum states can be verified, for example, by the violation of Bell's inequalities \cite{bell1964einstein,clauser1969proposed,PhysRevLett.28.938,PhysRevLett.49.91,PhysRevLett.115.250401}, quantum state tomography \cite{James2001}, entanglement witnesses \cite{PhysRevA.66.062305,barbieri2003detection,Bertlmann_PRA_witness-2005,PhysRevLett.105.230404,PhysRevLett.113.170402,PhysRevA.91.032315}, and measurements employing multiple copies of the quantum state \cite{Horodecki2003,Schmid2008,Mintert2007,PhysRevA.84.052112,zhang2013direct,islam2015}. For a general bipartite quantum state, all these methods require detection of both particles (subsystems). Known methods of verifying entanglement by performing measurement on one subsystem require the bipartite state to be pure \cite{walborn2006,Pires2009,Just2013,sharapova2015schmidt}. Whether the entanglement of a bipartite mixed state can be verified by detecting only one subsystem is a question of fundamental importance.
\par
We show by an example that it is indeed possible to verify the entanglement in a two-particle mixed state without detecting one of the particles. No coincidence measurement or postselection is required in our method. In order to demonstrate our method, we choose a polarization entangled mixed state, which can be obtained by generalizing two Bell states. We use two identical sources of the quantum state but only one pair of photons is produced at a time, i.e., multiple copies of the state are \emph{not} produced. We employ an interferometric technique to show that an entanglement criterion, namely the positive partial transpose
criterion \cite{Peres-sep-cond-PRL,Horodeckis-separability-PLA}, can be tested and the entanglement can also be measured by means of the concurrence, a popular measure of two-qubit entanglement \cite{hill1997entanglement,wootters1998entanglement}.
\par
Here, we present a detailed theoretical analysis of the entanglement verification technique. Furthermore, the nonlinear interferometer \cite{ZWM-ind-coh-PRL,WZM-ind-coh-PRA,chekhova2016nonlinear} used in our scheme has recently found important applications to various branches of quantum science and technology, including quantum imaging \cite{lemos2014,lahiri2015}, quantum spectroscopy \cite{kulik2016}, polarization control \cite{PhysRevA.95.033816}, and fundamental test of quantum mechanics \cite{Milonni_PRLindcoh,Milonni_PRAindcoh}. All the corresponding theoretical analyses apply to pure states only. Our theoretical analysis also shows how to treat a mixed state in such an interferometric arrangement. The experimental results are included in a separate publication \cite{pol-ent-exp}. 
\par
The article is organized as follows. In Sec. \ref{Sec:gen-scheme} we outline our entanglement analysis scheme. In Sec. \ref{sec:q-state-meas}, we present the class of states we address and discuss the relevant entanglement criterion and measure. In Sec. \ref{sec:ent-measurement}, we present the analysis of our method and the main results, which also include illustrations by numerical examples. Finally, in Sec. \ref{Sec:conc} we summarize and present our conclusions. 

\section{Outline of the Scheme}\label{Sec:gen-scheme}
\begin{figure*}\centering
  \includegraphics[width=0.8\textwidth]{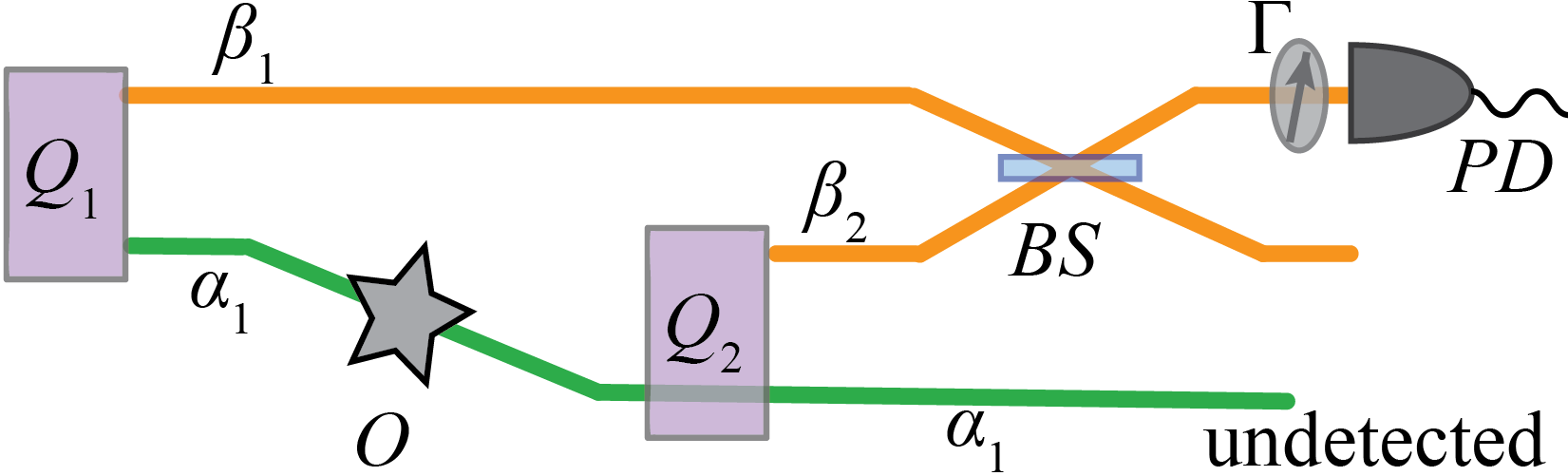}
\caption{Entanglement verification scheme. Two identical sources, $Q_1$ and $Q_2$, individually generate the same two-photon state ($\dm$). Source $Q_1$ can emit a photon pair ($\alpha$, $\beta$) into propagation modes $\alpha_{1}$ and $\beta_{1}$. Source $Q_2$ is restricted to emit photon $\a$ also in the mode $\a_1$. Photon $\alpha$, which is never detected, interacts with a device, $O$, between $Q_1$ and $Q_2$. Source $Q_2$ can emit photon $\b$ in propagation mode $\b_2$. Modes $\beta_1$ and $\beta_2$ are combined by a beamsplitter ($BS$) and an output of $BS$ is collected by a photo-detector ($PD$). Another device ($\Gamma$), placed before $PD$, allows us to choose the measurement basis. Sources $Q_1$ and $Q_2$ emit probabilistically and never emit simultaneously. When it is impossible to know the source of a detected photon, single-photon interference is observed at $PD$. For certain choices of basis, the entanglement of the two-photon state determines the visibility of the interference pattern. Information about the entanglement is retrieved from the single-photon interference patterns.} \label{fig:scheme-illus}
\end{figure*}
We consider two identical sources, $Q_1$ and $Q_2$, each of which can
produce the same two-photon mixed state, $\dm$; however, they never emit
simultaneously (Fig. \ref{fig:scheme-illus}). We denote the two
photons by $\a$ and $\b$. Suppose that $Q_1$ can emit photon $\a$ into propagation mode $\a_1$. We ensure that $Q_2$ can emit photon $\a$ only in the same propagation mode ($\a_1$). This is done by sending the beam of photon $\a$ generated by $Q_1$ through source $Q_2$ and perfectly aligning the beam with the beam of photon $\a$ generated by $Q_2$. Therefore, if one only observes photon $\a$ that emerges from $Q_2$, one cannot identify the origin of the photon.
\par
Sources $Q_1$ and $Q_1$ can emit photon $\b$ into distinct propagation modes $\b_1$ and $\b_2$, respectively. These two modes are superposed by a beam splitter, $BS$, and one of the outputs of $BS$
is directed to a photon detector, $PD$. A device, $\Gamma$, is
placed in front of $PD$ to choose appropriate measurement bases. It
is important to note that only the single-photon counting rate
(intensity) is measured in the experiment and no coincidence measurement is ever performed.
\par
A device, $O$, is placed in the path of photon $\a$ between $Q_1$ and $Q_2$. (This device does not affect the emission rates at $Q_1$ and $Q_2$.) Although $\a$ is never detected, the interaction with $O$ affects the interference pattern recorded by detecting photon $\b$ at $PD$
\cite{ZWM-ind-coh-PRL,WZM-ind-coh-PRA}. This striking phenomenon forms the basis of our entanglement verification scheme. We show that with the knowledge of this
interaction, the information about the entanglement in the two-photon quantum state can be retrieved from
single-photon interference patterns recorded in certain measurement bases. It is evident that the choice of devices $O$ and $\Gamma$ depends on the entangled degree of freedom.
\par
In order to illustrate the scheme we work with a two-photon
polarization entangled state which is discussed in the next section.

\section{The Quantum State}\label{sec:q-state-meas}
We consider a two-photon
polarization-entangled mixed state that can be characterized by
three free parameters. Such a state can be expressed in the general form \footnote{It is important to note that our entanglement verification scheme can also be applied to states $\dm=I_1 \ket{H_\alpha V_\beta }\bra{H_\alpha V_\beta }
+I_2 \ket{V_\alpha H_\beta }\bra{V_\alpha H_\beta }
+\big(\I\sqrt{I_1 I_2}e^{-i\phi}\ket{H_\alpha V_\beta }\bra{V_\alpha H_\beta }+\text{H.c.}\big)$.}
\begin{align}\label{mixed-state-form}
\dm=&I_H \ket{H_{\a},H_{\b}}\bra{H_{\a},H_{\b}} +I_V
\ket{V_{\a},V_{\b}}\bra{V_{\a},V_{\b}} \nonumber
\\& +\big(e^{-i\phi}\I\sqrt{I_H I_V}\ket{H_{\a},H_{\b}}
\bra{V_{\a},V_{\b}} +\text{H.c.}\big),
\end{align}
where $0 \leq I_H \leq 1$, $I_V=1-I_H$, $\phi$ is a
phase, $0\leq\I\leq 1$, and $H$ and $V$ represent horizontal and
vertical directions of polarization respectively. It is evident that $I_H$, $I_V$, $\I$, and $\phi$ are all real quantities. When $\I=1$, the
density operator ($\dm$) represents a pure state. When $\I=0$ and $I_H=1/2$ the state is diagonal in all bases, i.e. maximally mixed.
\par 
It is to be noted that $\dm$ can be obtained by generalizing the two following Bell States: $\ket{\Phi^{+}}=(\ket{H_{\a},H_{\b}}+\ket{V_{\a},V_{\b}})/\sqrt{2}$ and $\ket{\Phi^{-}}=(\ket{H_{\a},H_{\b}}-\ket{V_{\a},V_{\b}})/\sqrt{2}$.
\par
\emph{PPT Criterion}.\textemdash ~Since we have a bipartite two
dimensional entangled state, the criterion of positive partial
transpose (PPT criterion) \cite{Peres-sep-cond-PRL} can be applied
to ensure separability or entanglement
\cite{Horodeckis-separability-PLA}. A partial transposition of
two-particle density matrix ($\dm$) is a transposition taken with
respect to only one of the particles. The density operator $\dm$ has a positive partial transpose if and only if its partial
transposition does not have any negative eigenvalues. According to
the PPT criterion, a bipartite two-dimensional state is separable if
and only if $\dm$ has positive partial transpose
(\cite{Horodeckis-separability-PLA}, see also
\cite{GT-ent-det-Phrep-2009}).
\par
We take the partial transposition of $\dm$ [Eq.
(\ref{mixed-state-form})] with respect to photon $\a$ and find that
the resulting matrix has the following eigenvalues: \[ I_H,\quad
I_V, \quad \I\sqrt{I_H I_V}, \quad -\I\sqrt{I_H I_V}, \] where
$I_V=1-I_H$. Since $\I$, $I_H$, and $I_V$ cannot take negative
values, it follows from the PPT criterion that the state is
entangled if and only if
\begin{align}\label{PPT}
\I\sqrt{I_H I_V}\neq 0.
\end{align}
\par
\emph{Concurrence}. \textemdash ~The amount of entanglement in the
two-qubit state ($\dm$) can be quantified by the concurrence
\cite{wootters1998entanglement}. In order to determine the
concurrence one first needs to find the so-called ``spin-flipped''
density operator
\begin{align}\label{sf-rho}
\hat{\tilde{\rho}}=(\sy\otimes\sy)\dm^{\ast}(\sy\otimes\sy),
\end{align}
where $\sy$ is the second Pauli operator, $\otimes$ implies
Kronecker product, and the asterisk ($\ast$) refers to the complex
conjugation. The product $\dm\hat{\tilde{\rho}}$ has only real and
non-negative eigenvalues. If the square roots of these
eigenvalues, in decreasing order, are $\lambda_1$, $\lambda_2$,
$\lambda_3$, and $\lambda_4$, the concurrence of $\dm$ is given by
\begin{align}\label{conc-def}
\mathcal{C}(\dm)=\text{max}\{\lambda_1-\lambda_2-\lambda_3-\lambda_4,0\}.
\end{align}
\par
It follows from Eqs. (\ref{mixed-state-form}), (\ref{sf-rho}), and
(\ref{conc-def}) that in our case, the concurrence of the quantum
state is
\begin{align}\label{conc-form}
\mathcal{C}(\dm)=2\I\sqrt{I_H I_V}=2\I\sqrt{I_H (1-I_H)}.
\end{align}
\par
Below we show that for the quantum state given by Eq. (\ref{mixed-state-form}), the scheme allows us to test the PPT criterion as well as to measure the concurrence.

\section{Entanglement Verification and Measurement} \label{sec:ent-measurement}

\subsection{Physical Realization}\label{subsec:exp-details}
\begin{figure}[htbp] \centering
 \includegraphics[width=0.45\textwidth]{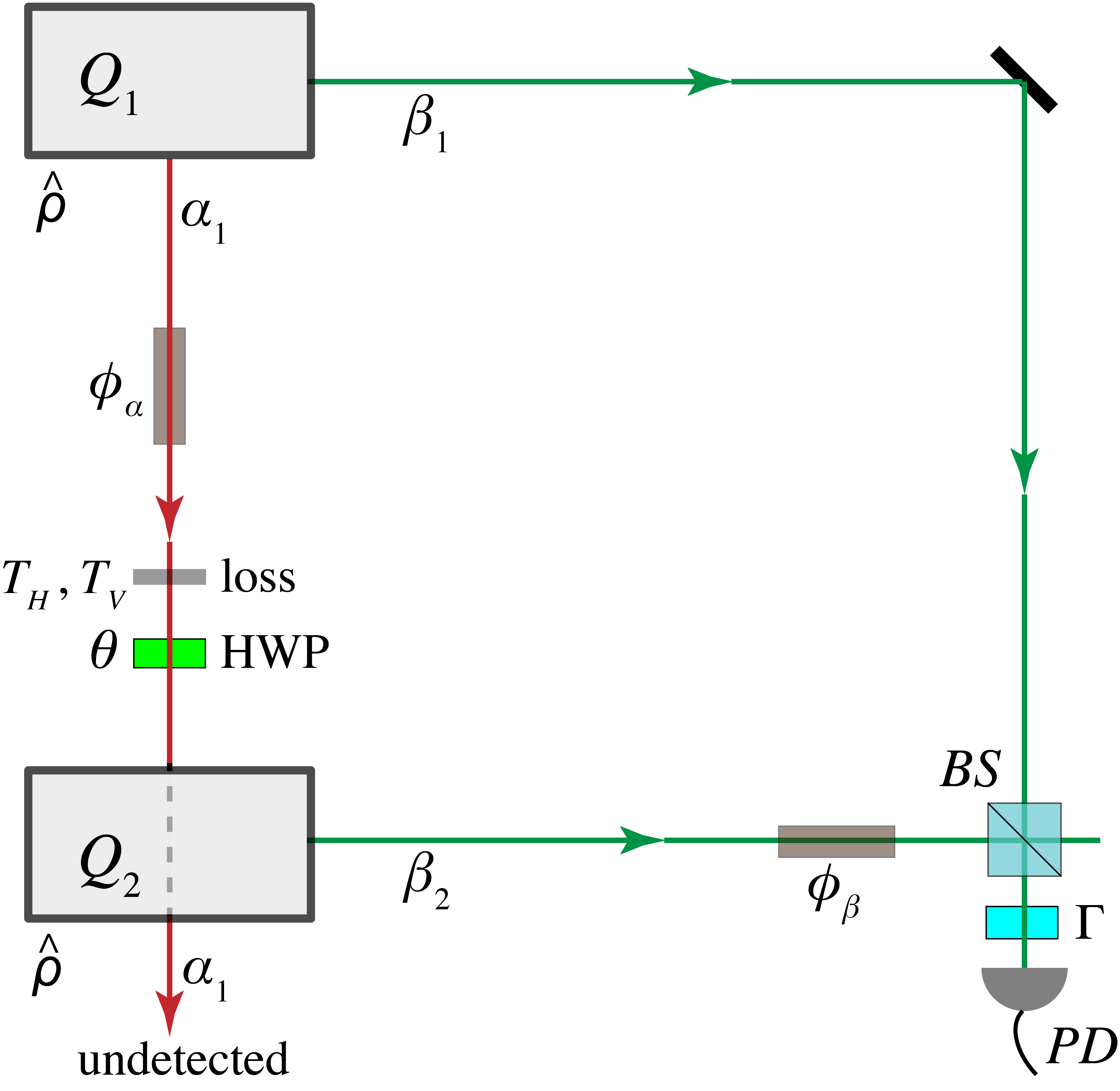}
  \qquad
\captionof{figure}{Entanglement verification of a polarization-entangled state. Each source ($Q_1$, $Q_2$) individually produces the state $\dm$ [Eq. (\ref{mixed-state-form})]. Device $O$ of Fig. \ref{fig:scheme-illus} is now a half-wave plate (HWP) and device $\Gamma$ projects photon $\b$ onto horizontal ($H$), vertical ($V$), diagonal ($D$), anti-diagonal ($A$), right-circular ($R$) or left-circular ($L$) polarization state. Photon $\b$ is detected at $PD$ and photon $\a$ is not detected. For certain choices of HWP angle and certain choices of the measurement basis, the single-photon interference pattern recorded at $PD$ contains information of entanglement in state $\dm$.} 
\label{fig:set-up} 
\end{figure}
The quantum state under consideration is entangled in polarization. In this case, we choose the device $O$ of Fig. \ref{fig:scheme-illus} to be a half-wave plate (HWP). As for device $\Gamma$, we use a combination of wave plates and a polarizer such that photon $\b$ can be projected onto the horizontal ($H$), vertical ($V$), diagonal ($D$), anti-diagonal ($A$), right-circular ($R$) or left-circular ($L$) polarization state. The experimental setup is illustrated in Fig. \ref{fig:set-up}.
\par
Below we provide a detailed theoretical analysis explaining how the information about entanglement can be obtained from the single-photon interference patterns recorded at $PD$.

\subsection{Deriving the Density Operator}\label{subsec:theory}
For the convenience of analysis, we rewrite Eq.
(\ref{mixed-state-form}) in the following form:
\begin{align}\label{mixed-state-j}
\dm_{j}=\sum_{\mu,\nu}\sqrt{I_{\mu}I_{\nu}} \I_{\mu\nu}
\exp(i\phi_{\mu\nu})
\ket{\mu_{\a}^{j},\mu_{\b}^{j}}\bra{\nu_{\a}^j,\nu_{\b}^j},
\end{align}
where $j=1,2$ refers to the sources $Q_1$ and $Q_2$; $\mu=H,V$;
$\nu=H,V$; $\phi_{\mu\nu}=0$ for $\mu=\nu$, and
$\phi_{HV}=-\phi_{VH}=-\phi$; and
\begin{equation}
\I_{\mu\nu}=\I_{\nu\mu}=\label{I-form}
\begin{cases}
1 & \quad \text{for} \quad \mu=\nu \\
\I & \quad \text{for} \quad \mu\neq \nu.
\end{cases}
\end{equation}
\par
We recall that sources produce only one photon pair in one detection run. The most general state that such an arrangement can produce is given by (Appendix 1)
\begin{align}\label{mixed-state-2source}
\dm_{\a\b}'=\sum_{j,k}^{1,2}\sum_{\mu,\nu}^{H,V}b_j b_k^{\ast}
\sqrt{I_{\mu}I_{\nu}}P_{\mu^j}^{\nu^k}\exp(i\xi_{\mu^j}^{\nu^k})
\ket{\mu_{\a}^{j},\mu_{\b}^{j}}\bra{\nu_{\a}^k,\nu_{\b}^k},
\end{align}
where both $j=1,2$ and $k=1,2$ refer to the sources, $|b_j|^2$ is the probability of the photon pair being emitted
by source $Q_j$, $|b_1|^2+|b_2|^2=1$; the asterisk ($\ast$) denotes
complex conjugation, the phase $\xi_{\mu^j}^{\nu^k}$ and the real
positive quantity $P_{\mu^j}^{\nu^k}$ must obey the
relations (Appendix 1)
\begin{subequations}\label{phase-P-form}
\begin{align}
&\xi_{\mu^j}^{\nu^k}=-\xi_{\nu^k}^{\mu^j}~~ \forall ~ j,k,\mu,\nu,
\quad \text{and} \quad
\xi_{\mu^j}^{\nu^j}=\phi_{\mu\nu}~~ \forall ~ j, \label{phase-P-form:a}\\
&P_{\mu^j}^{\nu^k}=P_{\nu^k}^{\mu^j}~~ \forall ~ j,k,\mu,\nu, \quad
\text{and} \quad P_{\mu^j}^{\nu^j}=\I_{\mu\nu}~~ \forall ~ j,
\label{phase-P-form:b}
\end{align}
\end{subequations}
and $0\leq P_{\mu^j}^{\nu^k}=P_{\nu^k}^{\mu^j} \leq 1$.
\par
We now impose the condition that  photon pairs emitted by separate
sources are fully coherent when they have the same polarization,
i.e.
\begin{align}\label{gen-coh-cond}
P_{\mu^1}^{\mu^2}=P_{\mu^2}^{\mu^1}=1 \quad \text{for}~~ \mu=H,V.
\end{align}
This condition is easily attained in the laboratory by employing
a pump laser with sufficiently long coherence length \cite{pol-ent-exp}. If we apply this condition, it follows from the
positive semi-definiteness of $\dm_{\a\b}'$ that (Appendix 2)
\begin{align}\label{P-I-rel}
P_{\mu^j}^{\nu^k}=\I_{\mu\nu},\quad \forall j,k,\mu,\nu,
\end{align}
where $\I_{\mu\nu}$ is given by Eq. (\ref{I-form}).
\par
From Eqs. (\ref{mixed-state-2source}) and (\ref{P-I-rel}), we find that the density operator
representing the state of a photon pair in our system is given by
\begin{align}\label{mixed-state-coh-em}
\dm_{\a\b}=\sum_{j,k}^{1,2}\sum_{\mu,\nu}^{H,V}b_j b_k^{\ast}
\sqrt{I_{\mu}I_{\nu}}\I_{\mu\nu}\exp(i\xi_{\mu^j}^{\nu^k})
\ket{\mu_{\a}^{j},\mu_{\b}^{j}}\bra{\nu_{\a}^k,\nu_{\b}^k}.
\end{align}
\par
We now mathematically represent the interaction of HWP with photon $\a$. Ideally, there should not be any loss of photon in the propagation mode $\a_1$ between $Q_1$ and $Q_2$. However, due to experimental imperfections it is almost impossible to avoid
slight misalignment of paths and probabilistic absorption of photon
$\a$ between $Q_1$ and $Q_2$. It order to make our measurement scheme robust against such losses, we take them into account quantitatively. Let $T_{\mu}$ be the probability
amplitude of photon $\a$ (polarized along direction $\mu$) to arrive
at $Q_2$ from $Q_1$. Without any loss of generality, we can assume
that $T_{\mu}$ is real and therefore $0<T_{\mu}<1$. The combined effect of HWP and the losses can be represented by the
following relations involving field operators:
\begin{subequations}\label{alignment-HWP-cond-1}
\begin{align}
\opa_{H_{\a}^{2}}&=e^{i\phi_{\a}}\left[T_H(\opa_{H_{\a}^{1}}\cos2\theta
+\opa_{V_{\a}^{1}}\sin2\theta)+R_H\opa_{H_{0}} \right],
\\
\opa_{V_{\a}^{2}}&=e^{i\phi_{\a}}\left[T_V(\opa_{H_{\a}^{1}}\sin2\theta
-\opa_{V_{\a}^{1}}\cos2\theta)+R_H\opa_{V_{0}} \right],
\end{align}
\end{subequations}
where $\opa$ represents photon annihilation operator:
$\opa_{\mu_{\a}^{j}}^{\dag}\ket{\text{vacuum}}=\ket{\mu_{\a}^{j}}$
and $\opa_{\mu_{0}}$ can be interpreted as the field of a lost
photon; $R_{\mu}=\sqrt{1-T_{\mu}^2}$; the HWP is set at angle $\theta$; and
$\phi_{\a}$ is the phase gained due to the propagation through air
from $Q_1$ to $Q_2$ (assumed to be the same for all polarization
directions). 
\par
We now proceed to derive the final form of the two-photon quantum state in the setup. Let us rewrite Eq. (\ref{alignment-HWP-cond-1}) in the
compact form
\begin{align}\label{alignment-HWP-cond}
\opa_{\mu_{\a}^{2}}=e^{i\phi_{\a}}\left[\sum_{\lambda}^{H,V}
\Lambda_{\mu\lambda}(\theta)\opa_{\lambda_{\a}^{1}}+
R_{\mu}\opa_{\mu_{0}}\right],\quad \mu=H,V,
\end{align}
where $\Lambda_{HH}(\theta)=T_H\cos2\theta$,
$\Lambda_{HV}(\theta)=T_H\sin2\theta$,
$\Lambda_{VH}(\theta)=T_V\sin2\theta$, and
$\Lambda_{VV}(\theta)=-T_V\cos2\theta$ are all real quantities.
Using the facts that
$\opa_{\mu_{\a}^{j}}^{\dag}\ket{\text{vacuum}}=\ket{\mu_{\a}^{j}}$
and $\opa_{\mu_{0}}^{\dag}\ket{\text{vacuum}}=\ket{\mu}_0$, we
obtain from Eq. (\ref{alignment-HWP-cond}) the following
transformation law for a ket:
\begin{align}\label{alignment-HWP-cond-st}
\ket{\mu_{\a}^{2}}=e^{-i\phi_{\a}}\left[\sum_{\lambda}^{H,V}
\Lambda_{\mu\lambda}(\theta)\ket{\lambda_{\a}^{1}}
+R_{\mu}\ket{\mu}_0\right],
\end{align}
where $\Lambda_{\mu\lambda}(\theta)$ is defined below Eq.
(\ref{alignment-HWP-cond}).
The final form of the density operator representing the two-photon quantum state is obtained by substituting Eq. (\ref{alignment-HWP-cond-st}) into
Eq. (\ref{mixed-state-coh-em}). We denote this density operator by $\dm_{\a\b}^{(f)}$ and provide its explicit form in Appendix 3.
\par
The density operator $\dm_{\a\b}^{(f)}$ (Appendix 3) can be used to determine the photon counting rate at the detector. Alternatively, one can
also use the reduced density operator ($\dm_{\b}$), which represents the state of photon $\b$ only. We use the latter approach in our analysis. We obtain the reduced density operator ($\dm_{\b}$) by taking partial trace of $\dm_{\a\b}^{(f)}$ over the subspace of photon $\a$ and the loss modes. We find it to have the form
\begin{align}\label{den-op-red}
&\dm_{\b}= |b_1|^2I_H\kbra{H_{\b}^{1}}{H_{\b}^{1}}+
|b_2|^2I_H\kbra{H_{\b}^{2}}{H_{\b}^{2}} \nonumber \\
&+|b_1|^2I_V\kbra{V_{\b}^{1}}{V_{\b}^{1}}+
|b_2|^2I_V\kbra{V_{\b}^{2}}{V_{\b}^{2}} \nonumber \\
&+ \bigg[b_1b_2^{\ast}\cos2\theta
\Big(I_HT_H\exp\{i(\phi_{\a}+\xi_{H^1}^{H^2})\}\kbra{H_{\b}^{1}}{H_{\b}^{2}}
\nonumber \\ & \qquad -
I_VT_V\exp\{i(\phi_{\a}+\xi_{V^1}^{V^2})\}\kbra{V_{\b}^{1}}{V_{\b}^{2}}\Big)+
\text{H.c.}\bigg] \nonumber \\ &+
\bigg[b_1b_2^{\ast}\I\sqrt{I_HI_V}\sin2\theta
\Big(T_V\exp\{i(\phi_{\a}+\xi_{H^1}^{V^2})\}\kbra{H_{\b}^{1}}{V_{\b}^{2}}
\nonumber \\ & \qquad +
T_H\exp\{i(\phi_{\a}+\xi_{V^1}^{H^2})\}\kbra{V_{\b}^{1}}{H_{\b}^{2}}\Big)+
\text{H.c.}\bigg].
\end{align}

\subsection{Determining Photon Counting Rates and Visibility}\label{subsec:phc-rt}
We now show how to determine the single-photon counting rate. We recall that the propagation modes $\b_1$ and $\b_2$ are
superposed by a beam splitter (BS) and one of the outputs of BS is sent
through a device, $\Gamma$, which
projects photon $\b$ onto a particular polarization state ($H$, $V$, $D$, $A$, $R$ or $L$). Therefore, the positive-frequency part of the quantized electric field at the detector can be represented by
\begin{align}\label{field-at-detector}
\opEps_{\mu_{\b}}=\opa_{\mu_{\b}^{1}}+ie^{i\phi_{\b}}\opa_{\mu_{\b}^{2}},
\quad \mu=H,V,D,A,R,L,
\end{align}
where $\opa_{\mu_{\b}^{j}}$ is the annihilation operator
corresponding to photon $\b$ with polarization $\mu$ in beam $\b_j$.
\par
The single-photon counting rate (for a given polarization) at the detector can now be obtained by the standard formula
\begin{align}\label{ph-count-rt}
\mathcal{R}_{\mu}=\avgr{\b}{\opEns_{\mu_{\b}}\opEps_{\mu_{\b}}},
\end{align}
where $\opEns_{\mu_{\b}}=\left\{ \opEps_{\mu_{\b}} \right\}^{\dag}$; and $\dm_{\b}$ and $\opEps_{\mu_{\b}}$ are given by Eqs. (\ref{den-op-red}) and (\ref{field-at-detector}) respectively. We show below in Sec. \ref{subsec:results} that the photon counting
rates measured for various polarizations represent various interference patterns. The visibility of such any such pattern is determined by the standard formula
\begin{align}\label{vis-def}
\mathcal{V}_{\mu}=\frac{\mathcal{R}^{\text{max}}_{\mu}
-\mathcal{R}^{\text{min}}_{\mu}} {\mathcal{R}^{\text{max}}_{\mu}
+\mathcal{R}^{\text{min}}_{\mu}},
\end{align}
where $\mu$ represents the polarization of the detected photon ($\b$), and
$\mathcal{R}^{\text{max}}_{\mu}$ and
$\mathcal{R}^{\text{min}}_{\mu}$ are, respectively, the maximum and
the minimum values of the single-photon counting rate.
\par
In the next section (Sec. \ref{subsec:results}), we show that the quantity $\I\sqrt{I_H I_V}$, which appears in the PPT-criterion [Eq.
(\ref{PPT})] and also in the formula of concurrence [Eq. (\ref{conc-form})], can be determined from the visibility of the above-mentioned  single-photon interference patterns.

\subsection{Signature of Entanglement in Single-photon Interference Patterns}\label{subsec:results}
We first consider the cases in which photon $\b$ is projected onto $\ket{H_\b}$ and $\ket{V_\b}$ polarization states. By the use of Eqs. (\ref{den-op-red}), (\ref{field-at-detector}), and (\ref{ph-count-rt}), we find that the photon counting rates are given by (we have applied
$|b_1|^2+|b_2|^2=1$)
\begin{subequations}\label{R-H/V}
\begin{align}
&\mathcal{R}_H=I_H \big\{1+2|b_1||b_2|T_H\cos2\theta
\sin(\phi_{in}+\xi_{H^1}^{H^2}) \big\}, \label{R-H/V:a}
\\&\mathcal{R}_V=I_V \big\{1-2|b_1||b_2|T_V\cos2\theta
\sin(\phi_{in}+\xi_{V^1}^{V^2}) \big\}, \label{R-H/V:b}
\end{align}
\end{subequations}
where
$\phi_{in}=\phi_{\a}-\phi_{\b}+\text{arg}\{b_1\}-\text{arg}\{b_2\}$ is the interferometric phase that is modulated to obtain the
interference patterns. It follows from Eqs. (\ref{vis-def}) and (\ref{R-H/V}) that
visibilities measured for $H$ and $V$ polarizations are
\begin{subequations}\label{vis-H/V}
\begin{align}
&\mathcal{V}_H= 2|b_1||b_2| T_H \cos2\theta, \label{vis-H}
\\&\mathcal{V}_V= 2|b_1||b_2| T_V \cos2\theta. \label{vis-V}
\end{align}
\end{subequations}
Clearly, when the half-wave plate is set at angle $\theta=0$, the visibilities measured for these polarizations have their maximum values
\begin{align}\label{vis-HV-0}
\mathcal{V}_H\big|_{\theta=0}= 2|b_1||b_2| T_H; \quad
\mathcal{V}_V\big|_{\theta=0}= 2|b_1||b_2| T_V.
\end{align}
We note that expressions of $\mathcal{V}_H$ and $\mathcal{V}_V$ do not contain $I_H$, $I_V$ or $\I$. Therefore, measurement in this basis does not yield any information about entanglement. However, $\mathcal{V}_H\big|_{\theta=0}$ and $\mathcal{V}_H\big|_{\theta=0}$ provide us with a quantitative measure of the photon loss in the propagation mode $\a_1$. Therefore, it is crucial to measure them in an actual experiment.  
\par
We now consider the case when the polarization of the detected
photon is diagonal ($D$), i.e., photon $\b$ is projected onto state $\ket{D_\b}$ \footnote{Note that $\ket{D}=(\ket{H}+\ket{V})/\sqrt{2}$.}. It follows from Eqs. (\ref{den-op-red}), (\ref{field-at-detector}), and
(\ref{ph-count-rt}) that the corresponding photon counting rate has the form
\begin{align}\label{R-D}
\mathcal{R}_D&=\frac{1}{2}\bigg[1+2|b_1||b_2|\cos2\theta \Big\{I_H
T_H \sin(\phi_{in}+\xi_{H^1}^{H^2})\nonumber
\\& \qquad \qquad \qquad \qquad \qquad-I_V T_V
\sin(\phi_{in}+\xi_{V^1}^{V^2})\Big\} \nonumber
\\&+2\I\sqrt{I_HI_V}|b_1||b_2|\sin2\theta \Big\{T_V
\sin(\phi_{in}+\xi_{H^1}^{V^2}) \nonumber
\\& \qquad \qquad \qquad \qquad \qquad +
T_H\sin(\phi_{in}+\xi_{V^1}^{H^2})\Big\} \bigg],
\end{align}
where we have applied the formulas $|b_1|^2+|b_2|^2=1$ and
$I_H+I_V=1$. We now set the half-wave plate angle to be $\pi/4$,
i.e., $\cos2\theta=0$ and $\sin2\theta=1$.
Under this condition Eq. (\ref{R-D}) reduces to \footnote{We have applied the trigonometric identity
$p\sin x+q\sin(x+\delta)=\sqrt{p^2+q^2+2pq\cos\delta}\sin(x+\zeta'')$, where $\tan\zeta''=q\sin\delta /(p+q\cos\delta)$.}
\begin{align}\label{R-D-90}
\mathcal{R}_D\big|_{\theta=\frac{\pi}{4}}=&\frac{1}{2}
\Big[1+2\I\sqrt{I_HI_V}|b_1||b_2|\nonumber
\\& \quad \times \sqrt{T_H^2+T_V^2+2T_HT_V\cos(\xi_{V^1}^{H^2}-\xi_{H^1}^{V^2})}\nonumber
\\& \qquad \qquad \qquad  \times
\sin(\phi_{in}+\zeta) \Big],
\end{align}
where
$\phi_{in}=\phi_{\a}-\phi_{\b}+\text{arg}\{b_1\}-\text{arg}\{b_2\}$, and $\tan\zeta=T_H\sin(\xi_{V^1}^{H^2}-\xi_{H^1}^{V^2})
/[T_V+T_H\cos(\xi_{V^1}^{H^2}-\xi_{H^1}^{V^2})]$; the explicit form of
$\zeta$ is not required for our purpose. 
\par
Following the same procedure, we find for $\ket{A_\b}$ that the photon counting rate is given by \footnote{Note that $\ket{A}=(\ket{V}-\ket{H})/\sqrt{2}$.}
\begin{align}\label{R-A-90}
\mathcal{R}_A\big|_{\theta=\frac{\pi}{4}}=&\frac{1}{2}
\Big[1-2\I\sqrt{I_HI_V}|b_1||b_2|\nonumber
\\& \quad \times \sqrt{T_H^2+T_V^2+2T_HT_V\cos(\xi_{V^1}^{H^2}-\xi_{H^1}^{V^2})}\nonumber
\\& \qquad \qquad \qquad  \times
\sin(\phi_{in}+\zeta) \Big].
\end{align}
\par
It follows from Eqs. (\ref{vis-def}), (\ref{R-D-90}), and
(\ref{R-A-90}) that the single-photon interference patterns recorded for $D$ and $A$ polarizations have the same visibility:
\begin{align}\label{vis-D-A-90}
\mathcal{V}_{D}\big|_{\theta=\frac{\pi}{4}}&=\mathcal{V}_{A}\big|_{\theta=\frac{\pi}{4}}
=2\I\sqrt{I_HI_V}|b_1||b_2|\nonumber
\\& \times
\sqrt{T_H^2+T_V^2+2T_HT_V\cos(\xi_{V^1}^{H^2}-\xi_{H^1}^{V^2})}.
\end{align}
We note that the concurrence ($\mathcal{C}(\dm)=2\I\sqrt{I_H I_V}$) appears in the formulas of $\mathcal{V}_{D}\big|_{\theta=\frac{\pi}{4}}$ and $\mathcal{V}_{A}\big|_{\theta=\frac{\pi}{4}}$. This fact implies that the single-photon interference patterns recorded for diagonal and anti-diagonal polarizations contain information about the amount of entanglement in the two-photon mixed state.
\par
Calculations for right-circular ($R$) and left-circular ($L$)
polarizations are very similar to those for diagonal and anti-diagonal
polarizations. When the half-wave plate angle is set such that
$\theta=\pi/4$, the corresponding photon counting rates become
\begin{subequations}\label{R-R-L-90}
\begin{align}
\mathcal{R}_R\big|_{\theta=\frac{\pi}{4}}=&\frac{1}{2}
\Big[1-2\I\sqrt{I_HI_V}|b_1||b_2|\nonumber
\\& \quad \times \sqrt{T_H^2+T_V^2-2T_HT_V\cos(\xi_{V^1}^{H^2}-\xi_{H^1}^{V^2})}\nonumber
\\& \qquad \qquad \qquad  \times
\sin(\phi_{in}-\zeta') \Big], \label{R-R-90} \\
\mathcal{R}_L\big|_{\theta=\frac{\pi}{4}}=&\frac{1}{2}
\Big[1+2\I\sqrt{I_HI_V}|b_1||b_2|\nonumber
\\& \quad \times \sqrt{T_H^2+T_V^2-2T_HT_V\cos(\xi_{V^1}^{H^2}-\xi_{H^1}^{V^2})}\nonumber
\\& \qquad \qquad \qquad  \times
\sin(\phi_{in}-\zeta') \Big], \label{R-L-90}
\end{align}
\end{subequations}
where
$\phi_{in}=\phi_{\a}-\phi_{\b}+\text{arg}\{b_1\}-\text{arg}\{b_2\}$
and $\tan\zeta'=T_H\sin(\xi_{V^1}^{H^2}-\xi_{H^1}^{V^2})
/[T_V-T_H\cos(\xi_{V^1}^{H^2}-\xi_{H^1}^{V^2})]$; the explicit form of
$\zeta'$ is not required for our purpose. Clearly, visibilities measured for $R$ and $L$ polarizations are given by
\begin{align}\label{vis-R-L-90}
\mathcal{V}_{R}\big|_{\theta=\frac{\pi}{4}}&=\mathcal{V}_{L}\big|_{\theta=\frac{\pi}{4}}=2\I\sqrt{I_HI_V}|b_1||b_2|\nonumber
\\& \times
\sqrt{T_H^2+T_V^2-2T_HT_V\cos(\xi_{V^1}^{H^2}-\xi_{H^1}^{V^2})}.
\end{align}
The presence of the concurrence ($\mathcal{C}(\dm)=2\I\sqrt{I_H I_V}$) in Eqs. (\ref{R-R-L-90}) and (\ref{vis-R-L-90}) shows that the single-photon interference patterns recorded for right-circular and left-circular polarizations contain information about the amount of entanglement in the two-photon mixed state.
\par
We note that the visibilities measured for diagonal, anti-diagonal, right-circular, and left-circular polarizations are linearly proportional to the concurrence of the two-photon state ($\dm$). 

\subsection{Test of the PPT Criterion}\label{subsec:PPT-test}
We now show that if $\I\sqrt{I_H I_V}\neq 0$, visibilities measured for diagonal ($D$) and right-circular ($R$) polarizations can never be simultaneously zero when the HWP-angle is set at $\theta=\pi/4$. It follows from Eqs. (\ref{vis-D-A-90}) and (\ref{vis-R-L-90}) that
\begin{align}\label{vis-PPT}
\left(\mathcal{V}_{D}\big|_{\theta=\frac{\pi}{4}}\right)^2
+\left(\mathcal{V}_{R}\big|_{\theta=\frac{\pi}{4}}\right)^2=& 8|b_1|^2|b_2|^2(T_H^2+T_V^2) \nonumber \\& \times (\I\sqrt{I_H I_V})^2.
\end{align}
Since $|b_1|$, $|b_2|$, $T_H$, and $T_V$ must be non-zero quantities, $\mathcal{V}_{D}\big|_{\theta=\frac{\pi}{4}}$ and $\mathcal{V}_{R}\big|_{\theta=\frac{\pi}{4}}$ can be simultaneously equal to zero if and only if $\I\sqrt{I_H I_V}= 0$. According to the PPT criterion (Sec. \ref{sec:q-state-meas}), the condition $\I\sqrt{I_H I_V}= 0$ implies that the two-photon mixed state [Eq. (\ref{mixed-state-form})] is separable. Since $\mathcal{V}_{D}\big|_{\theta=\frac{\pi}{4}}=\mathcal{V}_{A}\big|_{\theta=\frac{\pi}{4}}$ and $\mathcal{V}_{R}\big|_{\theta=\frac{\pi}{4}}=\mathcal{V}_{L}\big|_{\theta=\frac{\pi}{4}}$, it follows from the PPT criterion that when HWP is set at angle $\theta=\pi/4$, a non-zero value of the visibility (of the single-photon interference patterns) obtained for any one of polarizations $D$, $A$, $R$, and $L$ confirms that the two-photon mixed state, $\dm$ [Eq. (\ref{mixed-state-form})], is entangled. The state is separable (not entangled) if and only if visibilities measured for \emph{all of these polarizations} are zero. 
\par
We illustrate the test of PPT criterion by numerical examples in Sec. \ref{subsec:examples}.

\subsection{Determining the Concurrence}\label{subsec:conc-det}
It follows from the results of Sec. \ref{subsec:results} that the concurrence, $\mathcal{C}(\dm)$, of the two-photon mixed state can be determined from the single-photon patterns. By the use of Eqs. (\ref{conc-form}), (\ref{vis-HV-0}), and (\ref{vis-PPT}), we find that
\begin{align}\label{vis-comb-conc}
\mathcal{C}(\dm)=\sqrt{2
\frac{\left(\mathcal{V}_{D}\big|_{\theta=\frac{\pi}{4}}\right)^2
+\left(\mathcal{V}_{R}\big|_{\theta=\frac{\pi}{4}}\right)^2}
{\left(\mathcal{V}_{H}\big|_{\theta=0}\right)^2
+\left(\mathcal{V}_{V}\big|_{\theta=0}\right)^2}},
\end{align}
where $D$ and $R$ can be replaced by $A$ and $L$ respectively. \par
It follows from Eq. (\ref{vis-comb-conc}) that in order to determine the concurrence, one needs to measure visibilities not only for $\ket{D_\b}$ (or $\ket{A_\b}$) and $\ket{R_\b}$ (or $\ket{L_\b}$) but also for $\ket{H_\b}$ and $\ket{V_\b}$. However, we recall that although measurements corresponding to $\ket{D_\b}$, $\ket{A_\b}$, $\ket{R_\b}$, or $\ket{L_\b}$ yield information about entanglement, measurements corresponding to $\ket{H_\b}$ and $\ket{V_\b}$ do not. Therefore, it is natural to ask why measuring visibility for $\ket{H_\b}$ and $\ket{V_\b}$ is necessary to determine the concurrence. 
\par
Actually, under the ideal conditions ($|b_1|=|b_2|=1/\sqrt{2}$ and $T_H=T_V=1$), it is not required to measure the visibility for $\ket{H_\b}$ and $\ket{V_\b}$. It can be readily checked from Eq. (\ref{vis-HV-0}) that in this case the denominator on the right hand side of Eq. (\ref{vis-comb-conc}) is equal to $1$. However, no experimental situation is perfectly ideal. In particular, it is extremely challenging to achieve the condition $T_H=T_V=1$ due to photon losses and imperfect alignment. Furthermore, emission probabilities at the two sources ($Q_1$ and $Q_2$) may not be equal, i.e., the condition $|b_1|=|b_2|$ may not always apply. The measurement of visibility when photon $\b$ is projected onto states $\ket{H_\b}$ and $\ket{V_\b}$ allows us to take care of these experimental imperfections. In fact, Eq. (\ref{vis-comb-conc}) shows that in spite of all such imperfections being present, one is able to determine the concurrence by the use of our method. 
\par
We present the experimentally measured values of concurrence of different mixed states in a separate publication \cite{pol-ent-exp}. Below we provide some numerical examples to illustrate our results. 

\subsection{Numerical Illustration of Results}\label{subsec:examples}
We choose five density operators whose parameters are listed in Table \ref{table:qstate}. We determine the values of concurrence of these states by the use of Eq. (\ref{conc-form}) and find that states $\dm_1$ and $\dm_2$ are separable, whereas state $\dm_5$ is maximally entangled (a Bell state). Note that density operator $\dm_1$ represents a pure state and $\dm_2$ represents a fully mixed state. States $\dm_3$ and $\dm_4$ are neither maximally entangled nor separable.
\begin{center}
\begin{tabular}{| c | c | c | c | c |}
\hline
State & $I_H$ & $I_V$ & $\I$ & Concurrence \\
\hline \hline
$\dm_1$ & 1 & 0 & \textendash & 0 \\ [1.4pt]
$\dm_2$ & $0.5$ & $0.5$ & 0 & 0 \\ [1.4pt]
$\dm_3$ & $0.5$ & $0.5$ & 0.32 & 0.32 \\ [1.4pt]
$\dm_4$ & $0.5$ & $0.5$ & $0.5$ & $0.5$ \\ [1.4pt]
$\dm_5$ & $0.5$ & $0.5$ & $1$ & $1$ \\[1.4pt]
\hline
\end{tabular}
\captionof{table}{Two-photon mixed states used for illustration. The symbol ``\textendash'' implies not applicable. Parameter $\phi$ [see Eq. (\ref{mixed-state-form})] is not displayed because it plays no role in determining the amount of entanglement.}\label{table:qstate}
\end{center}
\par
For testing the PPT criterion with these states, we simulate an experimental situation in which experimental imperfections are present. In order to simulate the experimental imperfections, we assume that probabilities of emission at the two sources are not equal ($|b_1|\neq |b_2|$) and that there are photon losses in beam $\a_1$ due to imperfect alignment ($T_H \neq 1$, $T_V \neq 1$). The parameters are chosen as follows: $|b_1|^2=0.55$, $|b_2|^2=0.45$, $T_H=0.9$, $T_V=0.85$, and $\xi_{V^1}^{H^2}-\xi_{H^1}^{V^2}=\pi/4$. By the use of Eqs. (\ref{vis-D-A-90}) and (\ref{vis-R-L-90}), we compute the visibility of the single-photon patterns recorded for $\ket{D_\b}$, $\ket{A_\b}$, $\ket{R_\b}$, and $\ket{L_\b}$. The values of visibility are listed in Table \ref{table:PPT}.
\begin{center}
\begin{tabular}{| c | c | c | c | c | c |}
\hline
State & $\mathcal{V}_{D}\big|_{\theta=\frac{\pi}{4}}$ & $\mathcal{V}_{A}\big|_{\theta=\frac{\pi}{4}}$ & $\mathcal{V}_{R}\big|_{\theta=\frac{\pi}{4}}$ & $\mathcal{V}_{L}\big|_{\theta=\frac{\pi}{4}}$ & PPT criterion \\
\hline \hline
$\dm_1$ & $0$ & $0$ & $0$ & $0$ & separable \\ [1.4pt]
$\dm_2$ & $0$ & $0$ & $0$ & $0$ & separable \\ [1.4pt]
$\dm_3$ & $0.76$ & $0.76$ & $0.31$ & $0.31$ & entangled \\ [1.4pt]
$\dm_4$ & $0.40$ & $0.40$ & $0.17$ & $0.17$ & entangled \\ [1.4pt]
$\dm_5$ & $0.80$ & $0.80$ & $0.33$ & $0.33$ & entangled \\ [1.4pt]
\hline
\end{tabular}
\captionof{table}{Test of the PPT criterion for five different two-photon mixed states (Table \ref{table:qstate}). Choice of parameters: $|b_1|^2=0.55$, $|b_2|^2=0.45$, $T_H=0.9$, $T_V=0.85$, and $\xi_{V^1}^{H^2}-\xi_{H^1}^{V^2}=\pi/4$. For separable (not-entangled) states, visibilities measured for $\ket{D_\b}$, $\ket{A_\b}$, $\ket{R_\b}$, or $\ket{L_\b}$ are all zero.}\label{table:PPT}
\end{center}
\par
We find that for separable states ($\dm_1$ and $\dm_2$) the visibilities obtained for $\ket{D_\b}$, $\ket{A_\b}$, $\ket{R_\b}$, and $\ket{L_\b}$ are all equal to zero. Non-zero values of visibility for one of these polarizations confirm entanglement in the two-photon state.
\par
We now illustrate how the concurrence of the two-photon mixed state can be determined from the single-photon interference patterns even when experimental imperfections are present. For simplicity of notation, we denote the quantity $\sqrt{(\mathcal{V}_{D}\big|_{\theta=\frac{\pi}{4}})^2+(\mathcal{V}_{R}\big|_{\theta=\frac{\pi}{4}})^2}$ by $S$ and the quantity $\sqrt{[(\mathcal{V}_{H}\big|_{\theta=0})^2+(\mathcal{V}_{V}\big|_{\theta=0})^2]/2}$ by $N$. In this notation, the right hand side of Eq. (\ref{vis-comb-conc}) becomes $S/N$.
\begin{figure}[htbp]
\centering\includegraphics[width=0.9\linewidth]{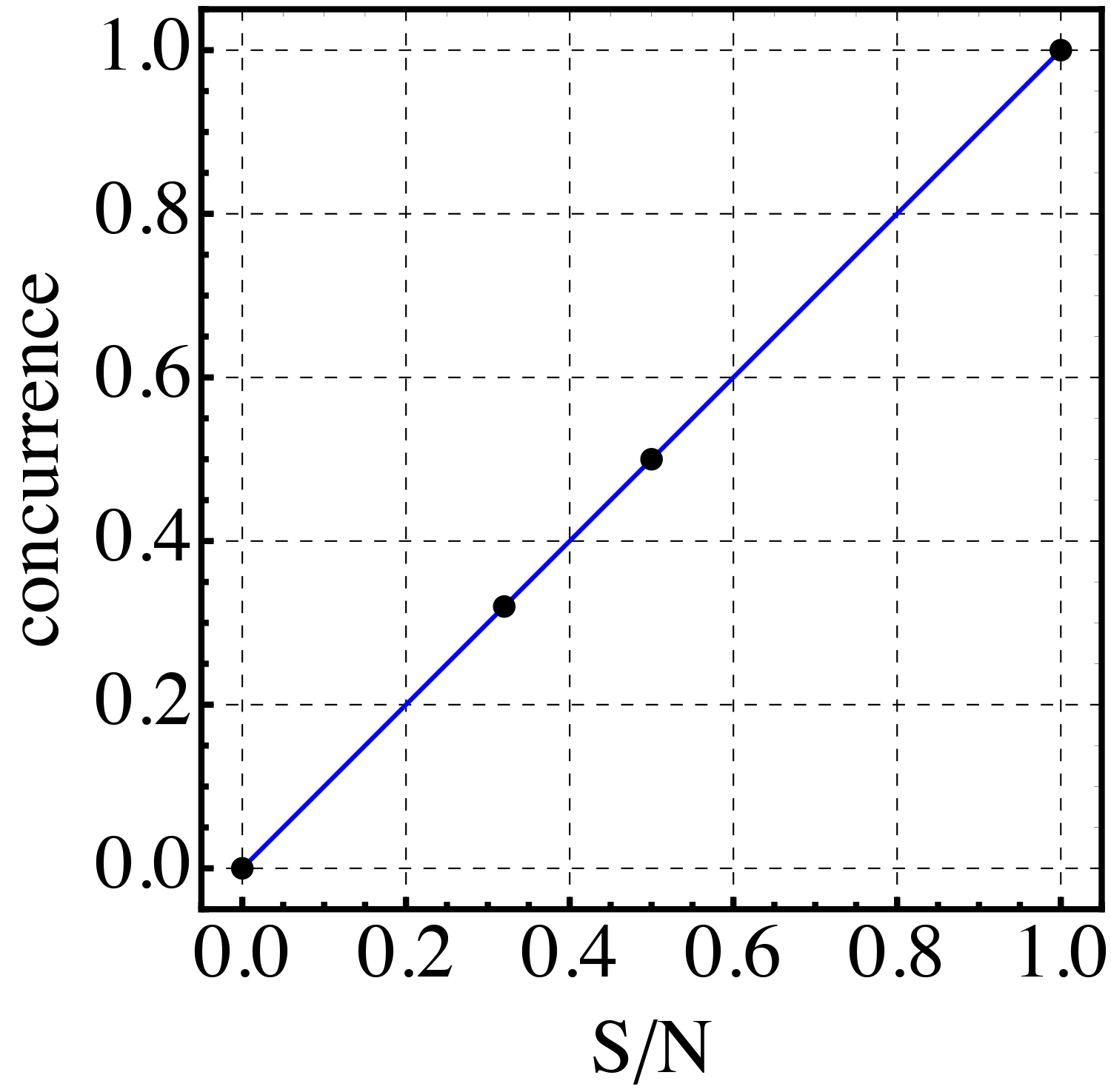}
\caption{Determining Concurrence from single-photon visibility. Experimental imperfections are simulated by choosing $|b_1|^2=0.55$, $|b_2|^2=0.45$, $T_H=0.9$, and $T_V=0.85$. Simulated data points (filled circles) represent computed values of the concurrence and of $S/N$ for five quantum states given by Table \ref{table:qstate}. (Data points for $\dm_1$ and $\dm_2$ coincide.) All simulated data points lie on the straight line predicted by Eq. (\ref{vis-comb-conc}) showing that the concurrence is equal to $S/N$ despite the presence of experimental imperfections.}\label{fig:concvisplot}
\end{figure}
\par
We choose the same experimental parameters given above. For the five states given by Table \ref{table:qstate}, we compute the values of the concurrence in two ways: i) by the use of Eq. (\ref{conc-form}); and ii) by determining the values of $S$ and $N$ from Eqs. (\ref{vis-HV-0}), (\ref{vis-D-A-90}), and (\ref{vis-R-L-90}).
In Fig. \ref{fig:concvisplot}, we plot the obtained values of the concurrence against $S/N$ and find that they lie exactly on the straight line predicted by Eq. (\ref{vis-comb-conc}). Since both $S$ and $N$ can be measured experimentally, Fig. \ref{fig:concvisplot} illustrates that the concurrence of a two-photon mixed state can be experimentally determined from single-photon interference patterns. 

\section{Summary and Conclusions}\label{Sec:conc}
We have shown that it is possible to verify and measure entanglement in a two-particle mixed state without detecting one of the particles and without any postselection. 
\par
To demonstrate our method, we have chosen a mixed state that can be obtained by generalizing two Bell states (see Sec. \ref{sec:q-state-meas}). It is straightforward to show that our method also applies to the mixed state which can be obtained by generalizing the other two Bell states \footnote{This mixed state has the form $\dm=I_1 \ket{H_{\a},V_{\b}}\bra{H_{\a},V_{\b}} +I_2
\ket{V_{\a},H_{\b}}\bra{V_{\a},H_{\b}}+(e^{-i\phi}\I\sqrt{I_1 I_2}\ket{H_{\a},V_{\b}}
\bra{V_{\a},H_{\b}} +\text{H.c.})$. It is obtained by generalizing the two Bell states $\ket{\Psi^{+}}=(\ket{H_{\a},V_{\b}}+\ket{V_{\a},H_{\b}})/\sqrt{2}$ and $\ket{\Psi^{-}}=(\ket{H_{\a},V_{\b}}-\ket{V_{\a},H_{\b}})/\sqrt{2}$.}. Therefore, our method currently covers all four Bell states and any two-dimensional mixed state that is obtained by generalizing them.
\par
Our method is based on the concept of path identity \cite{ZWM-ind-coh-PRL,WZM-ind-coh-PRA}. This concept has recently drawn considerable attention after it has found important applications to imaging \cite{lemos2014,lahiri2015}, spectroscopy \cite{kulik2016}, microwave superconducting cavities \cite{lahteenmaki2016coherence}, polarization control \cite{PhysRevA.95.033816}, optical coherence tomography \cite{valles2018optical,Paterova_2018}, the measurement of momentum correlation \cite{lahiri2017twin,hochrainer2017quantifying}, the generation of entangled states \cite{lahiri2018,MarioPhysRevLett2017,kysela2019experimental,Paraoanu_many-part-ent_PRA,zhang2012way}, and fundamental test of quantum mechanics \cite{Milonni_PRLindcoh,Milonni_PRAindcoh}. Like many of these applications, our entanglement verification method requires detection of only one of the particles of a two-particle system. Therefore, our method will be practically useful to test entanglement of a two-particle mixed state when a detector for one of the particles is not available. 
\par
Finally, our results open up a distinct avenue in verifying and measuring entanglement. They also inspire further questions. For example, one may now ask how to generalize the method so that it applies to many-particle high-dimensional entangled states. 

\section{Acknowledgements}\label{sec:ack}
We acknowledge support from the Austrian Academy of Sciences (\"OAW- 462 IQOQI, Vienna) and the Austrian Science Fund (FWF) with SFB F40 (FOQUS) and W1210-2 (CoQus). M.L. also acknowledges support from the College of Arts and Sciences and the Office of the Vice President for Research, Oklahoma State University. R.L. was supported by the National Science Centre (Poland) grants 2015/17/D/ST2/03471, 2015/16/S/ST2/00424, the Polish Ministry of Science and Higher Education, and the Foundation for Polish Science (FNP) under the FIRST TEAM project ``Spatiotemporal photon correlation measurements for quantum metrology and super-resolution microscopy'' cofinanced by the European Union under the European Regional Development Fund.

\section*{Appendix 1: Density operator represented by Eqs. (\ref{mixed-state-2source}) and (\ref{phase-P-form})}
\renewcommand{\theequation}{A-\arabic{equation}}

Here, we derive the density operator given by Eqs.
(\ref{mixed-state-2source}) and (\ref{phase-P-form}). 
\par
In order to derive the density operator, we first recall two important facts: 1) the sources $Q_1$ and $Q_2$ emit in such a way that they jointly produce only one photon pair at a time, i.e., the density operator must represent a state that is occupied by only two photons; and 2) the sources cannot produce biphoton states of the form $\ket{H_{\a},V_{\b}}$ and $\ket{V_{\a},H_{\b}}$. Therefore, the most general form that the density operator can take is given by
\begin{align}\label{mixed-state-2source-app}
\dm_{\a\b}'=\sum_{j,k}^{1,2}\sum_{\mu,\nu}^{H,V}b_j b_k^{\ast}
\sqrt{I_{\mu}^jI_{\nu}^k}P_{\mu^j}^{\nu^k}\exp(i\xi_{\mu^j}^{\nu^k})
\ket{\mu_{\a}^{j},\mu_{\b}^{j}}\bra{\nu_{\a}^k,\nu_{\b}^k},
\end{align}
where $j=1,2$ and $k=1,2$ represent the sources; $\mu=H,V$ and $\nu=H,V$; $|b_j|^2$ is the probability of the
photon pair being emitted by source $Q_j$; $|b_1|^2+|b_2|^2=1$; $I_{\mu}^j$ is the
probability with which source $Q_j$ emits the photon pair
$\ket{\mu_{\a},\mu_{\b}}$; $P_{\mu^j}^{\nu^k}$ is a non-negative real quantity; and $\xi_{\mu^j}^{\nu^k}$ is a phase (real quantity). Since $Q_1$ and $Q_2$ are identical sources, $I_{\mu}^j$ does not depend of $j$; we therefore drop this superscript and obtain the form (Eq.
(\ref{mixed-state-2source}) in main text)
\begin{align}\label{mixed-state-2source-appen}
\dm_{\a\b}'=\sum_{j,k}^{1,2}\sum_{\mu,\nu}^{H,V}b_j b_k^{\ast}
\sqrt{I_{\mu}I_{\nu}}P_{\mu^j}^{\nu^k}\exp(i\xi_{\mu^j}^{\nu^k})
\ket{\mu_{\a}^{j},\mu_{\b}^{j}}\bra{\nu_{\a}^k,\nu_{\b}^k},
\end{align}
One can readily check that this density operator has unit trace. Below we derive the conditions that the coefficients associated with the density operator must obey. 
\par
If $Q_2$ does not emit (i.e., $|b_2|=0$ and $|b_1|=1$), the density operator, $\dm_{\a\b}'$, must reduce to the state of light generated by $Q_1$ alone (i.e., state $\dm_1$ given by Eq. (\ref{mixed-state-j})). Likewise, if $Q_1$ does not emit (i.e., $|b_1|=0$ and $|b_2|=1$), the density operator, $\dm_{\a\b}'$, must reduce to the state of light generated by $Q_2$ alone (i.e., state $\dm_2$ given by Eq. (\ref{mixed-state-j})). Using these two facts, we immediately obtain
\begin{align}\label{app-1-res-1}
\xi_{\mu^j}^{\nu^j}=\phi_{\mu\nu}, \quad \text{and}
\quad P_{\mu^j}^{\nu^j}=\I_{\mu\nu}; \quad j=1,2.
\end{align}
Furthermore,
the density operator must be Hermitian. We therefore have 
\begin{align}\label{app-1-res-2}
\xi_{\mu^j}^{\nu^k}=-\xi_{\nu^k}^{\mu^j}, \quad \text{and} \quad
P_{\mu^j}^{\nu^k}=P_{\nu^k}^{\mu^j};~~ \forall ~ j,k,\mu,\nu.
\end{align}
This completes the derivation of the density operator given by Eqs. (\ref{mixed-state-2source}) and (\ref{phase-P-form}). 

\section*{Appendix 2: Derivation of Eq. (\ref{P-I-rel})}
In this appendix, we derive Eq. (\ref{P-I-rel}) by the use of Eqs. (\ref{mixed-state-2source}), (\ref{phase-P-form}), and (\ref{gen-coh-cond}). Eqs. (\ref{mixed-state-2source}) and (\ref{phase-P-form}) has already been re-displayed in Appendix 1. For the convenience of readers we display Eq. (\ref{gen-coh-cond}) once again below:
\begin{align}\label{gen-coh-cond-app}
P_{\mu^1}^{\mu^2}=P_{\mu^2}^{\mu^1}=1 \quad \text{for}~~ \mu=H,V.
\end{align}
\par
We note that the matrix elements of the density operator [Eq. (\ref{mixed-state-2source})]
contain four kinds of parameters: $b_j$, $I_{\mu}$,
$P_{\mu^j}^{\nu^k}$, and $\xi_{\mu^j}^{\nu^k}$. Each of them has a
distinct physical meaning, and their values do \emph{not} depend on each
other. For example, the value of $P_{\mu^j}^{\nu^k}$ does not change
if the values of $b_j$, $I_{\mu}$, and $\xi_{\mu^j}^{\nu^k}$ are changed.
This fact allows for a very simple derivation of Eq. (\ref{P-I-rel}). We
set $b_1=b_2=1/\sqrt{2}$, $I_H=I_V=1/2$, and
$\xi_{\mu^j}^{\nu^k}=0$ and represent the density operator in the following matrix form by combining Eqs. (\ref{mixed-state-2source}), (\ref{phase-P-form}), and (\ref{gen-coh-cond}):
\begin{align}\label{den-mat-sp-form}
\begin{pmatrix}
\frac{1}{4} & 0 & 0 & \frac{\I}{4} & \frac{1}{4} & 0 & 0 &
\frac{P_{H^1}^{V^2}}{4} \\
0 & 0 & 0 & 0 & 0 & 0 & 0 & 0 \\
0 & 0 & 0 & 0 & 0 & 0 & 0 & 0 \\
\frac{\I}{4} & 0 & 0 & \frac{1}{4} & \frac{P_{V^1}^{H^2}}{4} & 0 & 0
& \frac{1}{4} \\
\frac{1}{4} & 0 & 0 & \frac{P_{V^1}^{H^2}}{4} & \frac{1}{4} & 0 & 0
&
\frac{\I}{4} \\
0 & 0 & 0 & 0 & 0 & 0 & 0 & 0 \\
0 & 0 & 0 & 0 & 0 & 0 & 0 & 0 \\
\frac{P_{H^1}^{V^2}}{4} & 0 & 0 & \frac{1}{4} & \frac{\I}{4} & 0 & 0
& \frac{1}{4}
\end{pmatrix}.
\end{align}
Since a density matrix must be positive semi-definite, the real symmetric matrix given by Eq. (\ref{den-mat-sp-form}) must have nonnegative eigenvalues. The characteristic equation of this matrix can be expressed in the
following form:
\begin{align}\label{ch-eq}
x^8-x^7+c_2x^{6}-c_3x^{5}+c_4x^{4}=0,
\end{align}
the coefficients $c_2$, $c_3$, and $c_4$ are real numbers and the
solutions of this equation are the eigenvalues of the matrix. All
solutions of the characteristic equations are nonnegative when $c_j\geq 0$ for $j=2,3,4$. From the relation $c_3\geq 0$, we
obtain
\begin{align}\label{P-I-rel-interm}
\left(P_{H^1}^{V^2}-\I\right)^2+\left(P_{V^1}^{H^2}-\I\right)^2\leq 0.
\end{align}
Since the quantity on the left hand side of this inequality cannot take negative values, we must have
\begin{align}\label{P-I-rel-sp-app}
P_{H^1}^{V^2}=P_{V^1}^{H^2}=\I.
\end{align}
This relationships given by Eqs.  (\ref{gen-coh-cond}),
(\ref{phase-P-form:b}), and (\ref{P-I-rel-sp-app}) are jointly represented by Eq. (\ref{P-I-rel}):
\begin{align}\label{app-2-res-final}
P_{\mu^j}^{\nu^k}=\I_{\mu\nu},\quad \forall j,k,\mu,\nu.
\end{align}

\section*{Appendix 3: Explicit Form of the Final Density Operator}
Here we provide the explicit form of the density operator,
$\dm_{\a\b}^{(f)}$, representing the photon pair in our system. We
substitute Eq. (\ref{alignment-HWP-cond-st}) into Eq.
(\ref{mixed-state-coh-em}) and find that
\begin{widetext}
\begin{align}\label{mixed-state-coh-em-app}
\dm_{\a\b}^{(f)}&=\sum_{\mu,\nu}^{H,V}|b_1|^2
\sqrt{I_{\mu}I_{\nu}}\I_{\mu\nu}\exp(i\phi_{\mu\nu})
\ket{\mu_{\a}^{1},\mu_{\b}^{1}}\bra{\nu_{\a}^1,\nu_{\b}^1}+
\sum_{\mu,\nu}^{H,V}b_1b_2^{\ast}
\sqrt{I_{\mu}I_{\nu}}\I_{\mu\nu}\exp[i(\xi_{\mu^1}^{\nu^2}+\phi_{\a})]
R_{\nu}\ket{\mu_{\a}^{1},\mu_{\b}^{1}}\bra{\nu_{0},\nu_{\b}^2}
\nonumber
\\
&+ \sum_{\mu,\nu}^{H,V}b_1^{\ast}b_2
\sqrt{I_{\mu}I_{\nu}}\I_{\mu\nu}\exp[i(\xi_{\mu^2}^{\nu^1}-\phi_{\a})]
R_{\mu}\ket{\mu_{0},\mu_{\b}^2}\bra{\nu_{\a}^{1},\nu_{\b}^{1}}
\nonumber
\\
&+\sum_{\mu,\nu}^{H,V}|b_2|^2
\sqrt{I_{\mu}I_{\nu}}\I_{\mu\nu}\exp(i\phi_{\mu\nu})\Big[
R_{\mu}R_{\nu}\ket{\mu_{0},\mu_{\b}^{2}}\bra{\nu_{0},\nu_{\b}^2}+
R_{\nu}\sum_{\lambda}^{H,V}\Lambda_{\mu\lambda}(\theta)
\ket{\lambda_{\a}^{1},\mu_{\b}^{2}}\bra{\nu_{0},\nu_{\b}^2}
\nonumber
\\
& \qquad \qquad+
R_{\mu}\sum_{\lambda}^{H,V}\Lambda_{\nu\lambda}(\theta)
\ket{\mu_{0},\mu_{\b}^{2}}\bra{\lambda_{\a}^{1},\nu_{\b}^2}+
\sum_{\lambda}^{H,V}\sum_{\epsilon}^{H,V}\Lambda_{\mu\lambda}(\theta)
\Lambda_{\nu\epsilon}(\theta)
\ket{\lambda_{\a}^{1},\mu_{\b}^{2}}\bra{\epsilon_{\a}^1,\nu_{\b}^2}
\Big].
\end{align}
\end{widetext}

\bibliography{bib-Mandel-th-2019}

\begin{thebibliography}{48}
\expandafter\ifx\csname natexlab\endcsname\relax\def\natexlab#1{#1}\fi
\expandafter\ifx\csname bibnamefont\endcsname\relax
  \def\bibnamefont#1{#1}\fi
\expandafter\ifx\csname bibfnamefont\endcsname\relax
  \def\bibfnamefont#1{#1}\fi
\expandafter\ifx\csname citenamefont\endcsname\relax
  \def\citenamefont#1{#1}\fi
\expandafter\ifx\csname url\endcsname\relax
  \def\url#1{\texttt{#1}}\fi
\expandafter\ifx\csname urlprefix\endcsname\relax\def\urlprefix{URL }\fi
\providecommand{\bibinfo}[2]{#2}
\providecommand{\eprint}[2][]{\url{#2}}

\bibitem[{\citenamefont{G{\"u}hne and T{\'o}th}(2009)}]{GT-ent-det-Phrep-2009}
\bibinfo{author}{\bibfnamefont{O.}~\bibnamefont{G{\"u}hne}} \bibnamefont{and}
  \bibinfo{author}{\bibfnamefont{G.}~\bibnamefont{T{\'o}th}},
  \bibinfo{journal}{Physics Reports} \textbf{\bibinfo{volume}{474}},
  \bibinfo{pages}{1 } (\bibinfo{year}{2009}), ISSN \bibinfo{issn}{0370-1573},
  \urlprefix\url{http://www.sciencedirect.com/science/article/pii/S0370157309000623}.

\bibitem[{\citenamefont{Horodecki et~al.}(2009)\citenamefont{Horodecki,
  Horodecki, Horodecki, and Horodecki}}]{horodecki2009review}
\bibinfo{author}{\bibfnamefont{R.}~\bibnamefont{Horodecki}},
  \bibinfo{author}{\bibfnamefont{P.}~\bibnamefont{Horodecki}},
  \bibinfo{author}{\bibfnamefont{M.}~\bibnamefont{Horodecki}},
  \bibnamefont{and}
  \bibinfo{author}{\bibfnamefont{K.}~\bibnamefont{Horodecki}},
  \bibinfo{journal}{Rev. Mod. Phys.} \textbf{\bibinfo{volume}{81}},
  \bibinfo{pages}{865} (\bibinfo{year}{2009}),
  \urlprefix\url{https://link.aps.org/doi/10.1103/RevModPhys.81.865}.

\bibitem[{\citenamefont{Bell}(1964)}]{bell1964einstein}
\bibinfo{author}{\bibfnamefont{J.~S.} \bibnamefont{Bell}},
  \bibinfo{journal}{Physics} \textbf{\bibinfo{volume}{1}}, \bibinfo{pages}{195}
  (\bibinfo{year}{1964}).

\bibitem[{\citenamefont{Clauser et~al.}(1969)\citenamefont{Clauser, Horne,
  Shimony, and Holt}}]{clauser1969proposed}
\bibinfo{author}{\bibfnamefont{J.~F.} \bibnamefont{Clauser}},
  \bibinfo{author}{\bibfnamefont{M.~A.} \bibnamefont{Horne}},
  \bibinfo{author}{\bibfnamefont{A.}~\bibnamefont{Shimony}}, \bibnamefont{and}
  \bibinfo{author}{\bibfnamefont{R.~A.} \bibnamefont{Holt}},
  \bibinfo{journal}{Phys. Rev. Lett.} \textbf{\bibinfo{volume}{23}},
  \bibinfo{pages}{880} (\bibinfo{year}{1969}).

\bibitem[{\citenamefont{Freedman and Clauser}(1972)}]{PhysRevLett.28.938}
\bibinfo{author}{\bibfnamefont{S.~J.} \bibnamefont{Freedman}} \bibnamefont{and}
  \bibinfo{author}{\bibfnamefont{J.~F.} \bibnamefont{Clauser}},
  \bibinfo{journal}{Phys. Rev. Lett.} \textbf{\bibinfo{volume}{28}},
  \bibinfo{pages}{938} (\bibinfo{year}{1972}),
  \urlprefix\url{https://link.aps.org/doi/10.1103/PhysRevLett.28.938}.

\bibitem[{\citenamefont{Aspect et~al.}(1982)\citenamefont{Aspect, Grangier, and
  Roger}}]{PhysRevLett.49.91}
\bibinfo{author}{\bibfnamefont{A.}~\bibnamefont{Aspect}},
  \bibinfo{author}{\bibfnamefont{P.}~\bibnamefont{Grangier}}, \bibnamefont{and}
  \bibinfo{author}{\bibfnamefont{G.}~\bibnamefont{Roger}},
  \bibinfo{journal}{Phys. Rev. Lett.} \textbf{\bibinfo{volume}{49}},
  \bibinfo{pages}{91} (\bibinfo{year}{1982}),
  \urlprefix\url{https://link.aps.org/doi/10.1103/PhysRevLett.49.91}.

\bibitem[{\citenamefont{Giustina et~al.}(2015)\citenamefont{Giustina,
  Versteegh, Wengerowsky, Handsteiner, Hochrainer, Phelan, Steinlechner,
  Kofler, Larsson, Abell\'an et~al.}}]{PhysRevLett.115.250401}
\bibinfo{author}{\bibfnamefont{M.}~\bibnamefont{Giustina}},
  \bibinfo{author}{\bibfnamefont{M.~A.~M.} \bibnamefont{Versteegh}},
  \bibinfo{author}{\bibfnamefont{S.}~\bibnamefont{Wengerowsky}},
  \bibinfo{author}{\bibfnamefont{J.}~\bibnamefont{Handsteiner}},
  \bibinfo{author}{\bibfnamefont{A.}~\bibnamefont{Hochrainer}},
  \bibinfo{author}{\bibfnamefont{K.}~\bibnamefont{Phelan}},
  \bibinfo{author}{\bibfnamefont{F.}~\bibnamefont{Steinlechner}},
  \bibinfo{author}{\bibfnamefont{J.}~\bibnamefont{Kofler}},
  \bibinfo{author}{\bibfnamefont{J.-A.} \bibnamefont{Larsson}},
  \bibinfo{author}{\bibfnamefont{C.}~\bibnamefont{Abell\'an}},
  \bibnamefont{et~al.}, \bibinfo{journal}{Phys. Rev. Lett.}
  \textbf{\bibinfo{volume}{115}}, \bibinfo{pages}{250401}
  (\bibinfo{year}{2015}),
  \urlprefix\url{https://link.aps.org/doi/10.1103/PhysRevLett.115.250401}.

\bibitem[{\citenamefont{James et~al.}(2001)\citenamefont{James, Kwiat, Munro,
  and White}}]{James2001}
\bibinfo{author}{\bibfnamefont{D.~F.~V.} \bibnamefont{James}},
  \bibinfo{author}{\bibfnamefont{P.~G.} \bibnamefont{Kwiat}},
  \bibinfo{author}{\bibfnamefont{W.~J.} \bibnamefont{Munro}}, \bibnamefont{and}
  \bibinfo{author}{\bibfnamefont{A.~G.} \bibnamefont{White}},
  \bibinfo{journal}{Phys. Rev. A} \textbf{\bibinfo{volume}{64}},
  \bibinfo{pages}{052312} (\bibinfo{year}{2001}),
  \urlprefix\url{https://link.aps.org/doi/10.1103/PhysRevA.64.052312}.

\bibitem[{\citenamefont{G\"uhne et~al.}(2002)\citenamefont{G\"uhne, Hyllus,
  Bru\ss{}, Ekert, Lewenstein, Macchiavello, and Sanpera}}]{PhysRevA.66.062305}
\bibinfo{author}{\bibfnamefont{O.}~\bibnamefont{G\"uhne}},
  \bibinfo{author}{\bibfnamefont{P.}~\bibnamefont{Hyllus}},
  \bibinfo{author}{\bibfnamefont{D.}~\bibnamefont{Bru\ss{}}},
  \bibinfo{author}{\bibfnamefont{A.}~\bibnamefont{Ekert}},
  \bibinfo{author}{\bibfnamefont{M.}~\bibnamefont{Lewenstein}},
  \bibinfo{author}{\bibfnamefont{C.}~\bibnamefont{Macchiavello}},
  \bibnamefont{and} \bibinfo{author}{\bibfnamefont{A.}~\bibnamefont{Sanpera}},
  \bibinfo{journal}{Phys. Rev. A} \textbf{\bibinfo{volume}{66}},
  \bibinfo{pages}{062305} (\bibinfo{year}{2002}),
  \urlprefix\url{https://link.aps.org/doi/10.1103/PhysRevA.66.062305}.

\bibitem[{\citenamefont{Barbieri et~al.}(2003)\citenamefont{Barbieri,
  De~Martini, Di~Nepi, Mataloni, D'Ariano, and
  Macchiavello}}]{barbieri2003detection}
\bibinfo{author}{\bibfnamefont{M.}~\bibnamefont{Barbieri}},
  \bibinfo{author}{\bibfnamefont{F.}~\bibnamefont{De~Martini}},
  \bibinfo{author}{\bibfnamefont{G.}~\bibnamefont{Di~Nepi}},
  \bibinfo{author}{\bibfnamefont{P.}~\bibnamefont{Mataloni}},
  \bibinfo{author}{\bibfnamefont{G.~M.} \bibnamefont{D'Ariano}},
  \bibnamefont{and}
  \bibinfo{author}{\bibfnamefont{C.}~\bibnamefont{Macchiavello}},
  \bibinfo{journal}{Phys. Rev. Lett.} \textbf{\bibinfo{volume}{91}},
  \bibinfo{pages}{227901} (\bibinfo{year}{2003}),
  \urlprefix\url{https://link.aps.org/doi/10.1103/PhysRevLett.91.227901}.

\bibitem[{\citenamefont{Bertlmann et~al.}(2005)\citenamefont{Bertlmann,
  Durstberger, Hiesmayr, and Krammer}}]{Bertlmann_PRA_witness-2005}
\bibinfo{author}{\bibfnamefont{R.~A.} \bibnamefont{Bertlmann}},
  \bibinfo{author}{\bibfnamefont{K.}~\bibnamefont{Durstberger}},
  \bibinfo{author}{\bibfnamefont{B.~C.} \bibnamefont{Hiesmayr}},
  \bibnamefont{and} \bibinfo{author}{\bibfnamefont{P.}~\bibnamefont{Krammer}},
  \bibinfo{journal}{Phys. Rev. A} \textbf{\bibinfo{volume}{72}},
  \bibinfo{pages}{052331} (\bibinfo{year}{2005}),
  \urlprefix\url{https://link.aps.org/doi/10.1103/PhysRevA.72.052331}.

\bibitem[{\citenamefont{Park et~al.}(2010)\citenamefont{Park, Lee, Kim, Choi,
  and Sim}}]{PhysRevLett.105.230404}
\bibinfo{author}{\bibfnamefont{H.~S.} \bibnamefont{Park}},
  \bibinfo{author}{\bibfnamefont{S.-S.~B.} \bibnamefont{Lee}},
  \bibinfo{author}{\bibfnamefont{H.}~\bibnamefont{Kim}},
  \bibinfo{author}{\bibfnamefont{S.-K.} \bibnamefont{Choi}}, \bibnamefont{and}
  \bibinfo{author}{\bibfnamefont{H.-S.} \bibnamefont{Sim}},
  \bibinfo{journal}{Phys. Rev. Lett.} \textbf{\bibinfo{volume}{105}},
  \bibinfo{pages}{230404} (\bibinfo{year}{2010}),
  \urlprefix\url{https://link.aps.org/doi/10.1103/PhysRevLett.105.230404}.

\bibitem[{\citenamefont{Dai et~al.}(2014)\citenamefont{Dai, Len, Teo, Englert,
  and Krivitsky}}]{PhysRevLett.113.170402}
\bibinfo{author}{\bibfnamefont{J.}~\bibnamefont{Dai}},
  \bibinfo{author}{\bibfnamefont{Y.~L.} \bibnamefont{Len}},
  \bibinfo{author}{\bibfnamefont{Y.~S.} \bibnamefont{Teo}},
  \bibinfo{author}{\bibfnamefont{B.-G.} \bibnamefont{Englert}},
  \bibnamefont{and} \bibinfo{author}{\bibfnamefont{L.~A.}
  \bibnamefont{Krivitsky}}, \bibinfo{journal}{Phys. Rev. Lett.}
  \textbf{\bibinfo{volume}{113}}, \bibinfo{pages}{170402}
  (\bibinfo{year}{2014}),
  \urlprefix\url{https://link.aps.org/doi/10.1103/PhysRevLett.113.170402}.

\bibitem[{\citenamefont{Bartkiewicz et~al.}(2015)\citenamefont{Bartkiewicz,
  Horodecki, Lemr, Miranowicz, and \ifmmode~\dot{Z}\else
  \.{Z}\fi{}yczkowski}}]{PhysRevA.91.032315}
\bibinfo{author}{\bibfnamefont{K.}~\bibnamefont{Bartkiewicz}},
  \bibinfo{author}{\bibfnamefont{P.}~\bibnamefont{Horodecki}},
  \bibinfo{author}{\bibfnamefont{K.}~\bibnamefont{Lemr}},
  \bibinfo{author}{\bibfnamefont{A.}~\bibnamefont{Miranowicz}},
  \bibnamefont{and}
  \bibinfo{author}{\bibfnamefont{K.}~\bibnamefont{\ifmmode~\dot{Z}\else
  \.{Z}\fi{}yczkowski}}, \bibinfo{journal}{Phys. Rev. A}
  \textbf{\bibinfo{volume}{91}}, \bibinfo{pages}{032315}
  (\bibinfo{year}{2015}),
  \urlprefix\url{https://link.aps.org/doi/10.1103/PhysRevA.91.032315}.

\bibitem[{\citenamefont{Horodecki}(2003)}]{Horodecki2003}
\bibinfo{author}{\bibfnamefont{P.}~\bibnamefont{Horodecki}},
  \bibinfo{journal}{Phys. Rev. Lett.} \textbf{\bibinfo{volume}{90}},
  \bibinfo{pages}{167901} (\bibinfo{year}{2003}),
  \urlprefix\url{https://link.aps.org/doi/10.1103/PhysRevLett.90.167901}.

\bibitem[{\citenamefont{Schmid et~al.}(2008)\citenamefont{Schmid, Kiesel,
  Wieczorek, Weinfurter, Mintert, and Buchleitner}}]{Schmid2008}
\bibinfo{author}{\bibfnamefont{C.}~\bibnamefont{Schmid}},
  \bibinfo{author}{\bibfnamefont{N.}~\bibnamefont{Kiesel}},
  \bibinfo{author}{\bibfnamefont{W.}~\bibnamefont{Wieczorek}},
  \bibinfo{author}{\bibfnamefont{H.}~\bibnamefont{Weinfurter}},
  \bibinfo{author}{\bibfnamefont{F.}~\bibnamefont{Mintert}}, \bibnamefont{and}
  \bibinfo{author}{\bibfnamefont{A.}~\bibnamefont{Buchleitner}},
  \bibinfo{journal}{Phys. Rev. Lett.} \textbf{\bibinfo{volume}{101}},
  \bibinfo{pages}{260505} (\bibinfo{year}{2008}),
  \urlprefix\url{https://link.aps.org/doi/10.1103/PhysRevLett.101.260505}.

\bibitem[{\citenamefont{Mintert and Buchleitner}(2007)}]{Mintert2007}
\bibinfo{author}{\bibfnamefont{F.}~\bibnamefont{Mintert}} \bibnamefont{and}
  \bibinfo{author}{\bibfnamefont{A.}~\bibnamefont{Buchleitner}},
  \bibinfo{journal}{Phys. Rev. Lett.} \textbf{\bibinfo{volume}{98}},
  \bibinfo{pages}{140505} (\bibinfo{year}{2007}),
  \urlprefix\url{https://link.aps.org/doi/10.1103/PhysRevLett.98.140505}.

\bibitem[{\citenamefont{Zhang et~al.}(2011)\citenamefont{Zhang, Yu, Chen, and
  Oh}}]{PhysRevA.84.052112}
\bibinfo{author}{\bibfnamefont{C.}~\bibnamefont{Zhang}},
  \bibinfo{author}{\bibfnamefont{S.}~\bibnamefont{Yu}},
  \bibinfo{author}{\bibfnamefont{Q.}~\bibnamefont{Chen}}, \bibnamefont{and}
  \bibinfo{author}{\bibfnamefont{C.~H.} \bibnamefont{Oh}},
  \bibinfo{journal}{Phys. Rev. A} \textbf{\bibinfo{volume}{84}},
  \bibinfo{pages}{052112} (\bibinfo{year}{2011}),
  \urlprefix\url{https://link.aps.org/doi/10.1103/PhysRevA.84.052112}.

\bibitem[{\citenamefont{Zhang et~al.}(2013)\citenamefont{Zhang, Yang, Yang,
  Song, and Cao}}]{zhang2013direct}
\bibinfo{author}{\bibfnamefont{L.-H.} \bibnamefont{Zhang}},
  \bibinfo{author}{\bibfnamefont{Q.}~\bibnamefont{Yang}},
  \bibinfo{author}{\bibfnamefont{M.}~\bibnamefont{Yang}},
  \bibinfo{author}{\bibfnamefont{W.}~\bibnamefont{Song}}, \bibnamefont{and}
  \bibinfo{author}{\bibfnamefont{Z.-L.} \bibnamefont{Cao}},
  \bibinfo{journal}{Phys. Rev. A} \textbf{\bibinfo{volume}{88}},
  \bibinfo{pages}{062342} (\bibinfo{year}{2013}).

\bibitem[{\citenamefont{Islam et~al.}(2015)\citenamefont{Islam, Ma, Preiss,
  Tai, Lukin, Rispoli, and Greiner}}]{islam2015}
\bibinfo{author}{\bibfnamefont{R.}~\bibnamefont{Islam}},
  \bibinfo{author}{\bibfnamefont{R.}~\bibnamefont{Ma}},
  \bibinfo{author}{\bibfnamefont{P.~M.} \bibnamefont{Preiss}},
  \bibinfo{author}{\bibfnamefont{M.~E.} \bibnamefont{Tai}},
  \bibinfo{author}{\bibfnamefont{A.}~\bibnamefont{Lukin}},
  \bibinfo{author}{\bibfnamefont{M.}~\bibnamefont{Rispoli}}, \bibnamefont{and}
  \bibinfo{author}{\bibfnamefont{M.}~\bibnamefont{Greiner}},
  \bibinfo{journal}{Nature} \textbf{\bibinfo{volume}{528}}, \bibinfo{pages}{77}
  (\bibinfo{year}{2015}).

\bibitem[{\citenamefont{Walborn et~al.}(2006)\citenamefont{Walborn, Ribeiro,
  Davidovich, Mintert, and Buchleitner}}]{walborn2006}
\bibinfo{author}{\bibfnamefont{S.}~\bibnamefont{Walborn}},
  \bibinfo{author}{\bibfnamefont{P.~S.} \bibnamefont{Ribeiro}},
  \bibinfo{author}{\bibfnamefont{L.}~\bibnamefont{Davidovich}},
  \bibinfo{author}{\bibfnamefont{F.}~\bibnamefont{Mintert}}, \bibnamefont{and}
  \bibinfo{author}{\bibfnamefont{A.}~\bibnamefont{Buchleitner}},
  \bibinfo{journal}{Nature} \textbf{\bibinfo{volume}{440}},
  \bibinfo{pages}{1022} (\bibinfo{year}{2006}).

\bibitem[{\citenamefont{Di~Lorenzo~Pires
  et~al.}(2009)\citenamefont{Di~Lorenzo~Pires, Monken, and van
  Exter}}]{Pires2009}
\bibinfo{author}{\bibfnamefont{H.}~\bibnamefont{Di~Lorenzo~Pires}},
  \bibinfo{author}{\bibfnamefont{C.~H.} \bibnamefont{Monken}},
  \bibnamefont{and} \bibinfo{author}{\bibfnamefont{M.~P.} \bibnamefont{van
  Exter}}, \bibinfo{journal}{Phys. Rev. A} \textbf{\bibinfo{volume}{80}},
  \bibinfo{pages}{022307} (\bibinfo{year}{2009}),
  \urlprefix\url{https://link.aps.org/doi/10.1103/PhysRevA.80.022307}.

\bibitem[{\citenamefont{Just et~al.}(2013)\citenamefont{Just, Cavanna,
  Chekhova, and Leuchs}}]{Just2013}
\bibinfo{author}{\bibfnamefont{F.}~\bibnamefont{Just}},
  \bibinfo{author}{\bibfnamefont{A.}~\bibnamefont{Cavanna}},
  \bibinfo{author}{\bibfnamefont{M.~V.} \bibnamefont{Chekhova}},
  \bibnamefont{and} \bibinfo{author}{\bibfnamefont{G.}~\bibnamefont{Leuchs}},
  \bibinfo{journal}{New Journal of Physics} \textbf{\bibinfo{volume}{15}},
  \bibinfo{pages}{083015} (\bibinfo{year}{2013}).

\bibitem[{\citenamefont{Sharapova et~al.}(2015)\citenamefont{Sharapova,
  P\'erez, Tikhonova, and Chekhova}}]{sharapova2015schmidt}
\bibinfo{author}{\bibfnamefont{P.}~\bibnamefont{Sharapova}},
  \bibinfo{author}{\bibfnamefont{A.~M.} \bibnamefont{P\'erez}},
  \bibinfo{author}{\bibfnamefont{O.~V.} \bibnamefont{Tikhonova}},
  \bibnamefont{and} \bibinfo{author}{\bibfnamefont{M.~V.}
  \bibnamefont{Chekhova}}, \bibinfo{journal}{Phys. Rev. A}
  \textbf{\bibinfo{volume}{91}}, \bibinfo{pages}{043816}
  (\bibinfo{year}{2015}),
  \urlprefix\url{https://link.aps.org/doi/10.1103/PhysRevA.91.043816}.

\bibitem[{\citenamefont{Peres}(1996)}]{Peres-sep-cond-PRL}
\bibinfo{author}{\bibfnamefont{A.}~\bibnamefont{Peres}},
  \bibinfo{journal}{Phys. Rev. Lett.} \textbf{\bibinfo{volume}{77}},
  \bibinfo{pages}{1413} (\bibinfo{year}{1996}),
  \urlprefix\url{https://link.aps.org/doi/10.1103/PhysRevLett.77.1413}.

\bibitem[{\citenamefont{Horodecki}(1997)}]{Horodeckis-separability-PLA}
\bibinfo{author}{\bibfnamefont{P.}~\bibnamefont{Horodecki}},
  \bibinfo{journal}{Physics Letters A} \textbf{\bibinfo{volume}{232}},
  \bibinfo{pages}{333 } (\bibinfo{year}{1997}), ISSN \bibinfo{issn}{0375-9601},
  \urlprefix\url{http://www.sciencedirect.com/science/article/pii/S0375960197004167}.

\bibitem[{\citenamefont{Hill and Wootters}(1997)}]{hill1997entanglement}
\bibinfo{author}{\bibfnamefont{S.}~\bibnamefont{Hill}} \bibnamefont{and}
  \bibinfo{author}{\bibfnamefont{W.~K.} \bibnamefont{Wootters}},
  \bibinfo{journal}{Phys. Rev. Lett.} \textbf{\bibinfo{volume}{78}},
  \bibinfo{pages}{5022} (\bibinfo{year}{1997}).

\bibitem[{\citenamefont{Wootters}(1998)}]{wootters1998entanglement}
\bibinfo{author}{\bibfnamefont{W.~K.} \bibnamefont{Wootters}},
  \bibinfo{journal}{Phys. Rev. Lett.} \textbf{\bibinfo{volume}{80}},
  \bibinfo{pages}{2245} (\bibinfo{year}{1998}).

\bibitem[{\citenamefont{Zou et~al.}(1991)\citenamefont{Zou, Wang, and
  Mandel}}]{ZWM-ind-coh-PRL}
\bibinfo{author}{\bibfnamefont{X.~Y.} \bibnamefont{Zou}},
  \bibinfo{author}{\bibfnamefont{L.~J.} \bibnamefont{Wang}}, \bibnamefont{and}
  \bibinfo{author}{\bibfnamefont{L.}~\bibnamefont{Mandel}},
  \bibinfo{journal}{Phys. Rev. Lett.} \textbf{\bibinfo{volume}{67}},
  \bibinfo{pages}{318} (\bibinfo{year}{1991}),
  \urlprefix\url{https://link.aps.org/doi/10.1103/PhysRevLett.67.318}.

\bibitem[{\citenamefont{Wang et~al.}(1991)\citenamefont{Wang, Zou, and
  Mandel}}]{WZM-ind-coh-PRA}
\bibinfo{author}{\bibfnamefont{L.~J.} \bibnamefont{Wang}},
  \bibinfo{author}{\bibfnamefont{X.~Y.} \bibnamefont{Zou}}, \bibnamefont{and}
  \bibinfo{author}{\bibfnamefont{L.}~\bibnamefont{Mandel}},
  \bibinfo{journal}{Phys. Rev. A} \textbf{\bibinfo{volume}{44}},
  \bibinfo{pages}{4614} (\bibinfo{year}{1991}),
  \urlprefix\url{https://link.aps.org/doi/10.1103/PhysRevA.44.4614}.

\bibitem[{\citenamefont{Chekhova and Ou}(2016)}]{chekhova2016nonlinear}
\bibinfo{author}{\bibfnamefont{M.}~\bibnamefont{Chekhova}} \bibnamefont{and}
  \bibinfo{author}{\bibfnamefont{Z.}~\bibnamefont{Ou}},
  \bibinfo{journal}{Advances in Optics and Photonics}
  \textbf{\bibinfo{volume}{8}}, \bibinfo{pages}{104} (\bibinfo{year}{2016}).

\bibitem[{\citenamefont{Lemos et~al.}(2014)\citenamefont{Lemos, Borish, Cole,
  Ramelow, Lapkiewicz, and Zeilinger}}]{lemos2014}
\bibinfo{author}{\bibfnamefont{G.~B.} \bibnamefont{Lemos}},
  \bibinfo{author}{\bibfnamefont{V.}~\bibnamefont{Borish}},
  \bibinfo{author}{\bibfnamefont{G.~D.} \bibnamefont{Cole}},
  \bibinfo{author}{\bibfnamefont{S.}~\bibnamefont{Ramelow}},
  \bibinfo{author}{\bibfnamefont{R.}~\bibnamefont{Lapkiewicz}},
  \bibnamefont{and}
  \bibinfo{author}{\bibfnamefont{A.}~\bibnamefont{Zeilinger}},
  \bibinfo{journal}{Nature} \textbf{\bibinfo{volume}{512}},
  \bibinfo{pages}{409} (\bibinfo{year}{2014}).

\bibitem[{\citenamefont{Lahiri et~al.}(2015)\citenamefont{Lahiri, Lapkiewicz,
  Lemos, and Zeilinger}}]{lahiri2015}
\bibinfo{author}{\bibfnamefont{M.}~\bibnamefont{Lahiri}},
  \bibinfo{author}{\bibfnamefont{R.}~\bibnamefont{Lapkiewicz}},
  \bibinfo{author}{\bibfnamefont{G.~B.} \bibnamefont{Lemos}}, \bibnamefont{and}
  \bibinfo{author}{\bibfnamefont{A.}~\bibnamefont{Zeilinger}},
  \bibinfo{journal}{Phys. Rev. A} \textbf{\bibinfo{volume}{92}},
  \bibinfo{pages}{013832} (\bibinfo{year}{2015}).

\bibitem[{\citenamefont{Kalashnikov et~al.}(2016)\citenamefont{Kalashnikov,
  Paterova, Kulik, and Krivitsky}}]{kulik2016}
\bibinfo{author}{\bibfnamefont{D.~A.} \bibnamefont{Kalashnikov}},
  \bibinfo{author}{\bibfnamefont{A.~V.} \bibnamefont{Paterova}},
  \bibinfo{author}{\bibfnamefont{S.~P.} \bibnamefont{Kulik}}, \bibnamefont{and}
  \bibinfo{author}{\bibfnamefont{L.~A.} \bibnamefont{Krivitsky}},
  \bibinfo{journal}{Nature Photonics} \textbf{\bibinfo{volume}{10}},
  \bibinfo{pages}{98} (\bibinfo{year}{2016}).

\bibitem[{\citenamefont{Lahiri et~al.}(2017{\natexlab{a}})\citenamefont{Lahiri,
  Hochrainer, Lapkiewicz, Lemos, and Zeilinger}}]{PhysRevA.95.033816}
\bibinfo{author}{\bibfnamefont{M.}~\bibnamefont{Lahiri}},
  \bibinfo{author}{\bibfnamefont{A.}~\bibnamefont{Hochrainer}},
  \bibinfo{author}{\bibfnamefont{R.}~\bibnamefont{Lapkiewicz}},
  \bibinfo{author}{\bibfnamefont{G.~B.} \bibnamefont{Lemos}}, \bibnamefont{and}
  \bibinfo{author}{\bibfnamefont{A.}~\bibnamefont{Zeilinger}},
  \bibinfo{journal}{Phys. Rev. A} \textbf{\bibinfo{volume}{95}},
  \bibinfo{pages}{033816} (\bibinfo{year}{2017}{\natexlab{a}}),
  \urlprefix\url{https://link.aps.org/doi/10.1103/PhysRevA.95.033816}.

\bibitem[{\citenamefont{Heuer et~al.}(2015{\natexlab{a}})\citenamefont{Heuer,
  Menzel, and Milonni}}]{Milonni_PRLindcoh}
\bibinfo{author}{\bibfnamefont{A.}~\bibnamefont{Heuer}},
  \bibinfo{author}{\bibfnamefont{R.}~\bibnamefont{Menzel}}, \bibnamefont{and}
  \bibinfo{author}{\bibfnamefont{P.~W.} \bibnamefont{Milonni}},
  \bibinfo{journal}{Phys. Rev. Lett.} \textbf{\bibinfo{volume}{114}},
  \bibinfo{pages}{053601} (\bibinfo{year}{2015}{\natexlab{a}}),
  \urlprefix\url{https://link.aps.org/doi/10.1103/PhysRevLett.114.053601}.

\bibitem[{\citenamefont{Heuer et~al.}(2015{\natexlab{b}})\citenamefont{Heuer,
  Menzel, and Milonni}}]{Milonni_PRAindcoh}
\bibinfo{author}{\bibfnamefont{A.}~\bibnamefont{Heuer}},
  \bibinfo{author}{\bibfnamefont{R.}~\bibnamefont{Menzel}}, \bibnamefont{and}
  \bibinfo{author}{\bibfnamefont{P.~W.} \bibnamefont{Milonni}},
  \bibinfo{journal}{Phys. Rev. A} \textbf{\bibinfo{volume}{92}},
  \bibinfo{pages}{033834} (\bibinfo{year}{2015}{\natexlab{b}}),
  \urlprefix\url{https://link.aps.org/doi/10.1103/PhysRevA.92.033834}.

\bibitem[{\citenamefont{Lemos et~al.}(2019)\citenamefont{Lemos, Lapkiewicz,
  Hochrainer, Lahiri, and Zeilinger}}]{pol-ent-exp}
\bibinfo{author}{\bibfnamefont{G.~B.} \bibnamefont{Lemos}},
  \bibinfo{author}{\bibfnamefont{R.}~\bibnamefont{Lapkiewicz}},
  \bibinfo{author}{\bibfnamefont{A.}~\bibnamefont{Hochrainer}},
  \bibinfo{author}{\bibfnamefont{M.}~\bibnamefont{Lahiri}}, \bibnamefont{and}
  \bibinfo{author}{\bibfnamefont{A.}~\bibnamefont{Zeilinger}},
  \bibinfo{journal}{submitted to Phys. Rev. Lett.}  (\bibinfo{year}{2019}).

\bibitem[{\citenamefont{L{\"a}hteenm{\"a}ki
  et~al.}(2016)\citenamefont{L{\"a}hteenm{\"a}ki, Paraoanu, Hassel, and
  Hakonen}}]{lahteenmaki2016coherence}
\bibinfo{author}{\bibfnamefont{P.}~\bibnamefont{L{\"a}hteenm{\"a}ki}},
  \bibinfo{author}{\bibfnamefont{G.~S.} \bibnamefont{Paraoanu}},
  \bibinfo{author}{\bibfnamefont{J.}~\bibnamefont{Hassel}}, \bibnamefont{and}
  \bibinfo{author}{\bibfnamefont{P.~J.} \bibnamefont{Hakonen}},
  \bibinfo{journal}{Nature communications} \textbf{\bibinfo{volume}{7}},
  \bibinfo{pages}{1} (\bibinfo{year}{2016}).

\bibitem[{\citenamefont{Vall{\'e}s et~al.}(2018)\citenamefont{Vall{\'e}s,
  Jim{\'e}nez, Salazar-Serrano, and Torres}}]{valles2018optical}
\bibinfo{author}{\bibfnamefont{A.}~\bibnamefont{Vall{\'e}s}},
  \bibinfo{author}{\bibfnamefont{G.}~\bibnamefont{Jim{\'e}nez}},
  \bibinfo{author}{\bibfnamefont{L.~J.} \bibnamefont{Salazar-Serrano}},
  \bibnamefont{and} \bibinfo{author}{\bibfnamefont{J.~P.}
  \bibnamefont{Torres}}, \bibinfo{journal}{Phys. Rev. A}
  \textbf{\bibinfo{volume}{97}}, \bibinfo{pages}{023824}
  (\bibinfo{year}{2018}).

\bibitem[{\citenamefont{Paterova et~al.}(2018)\citenamefont{Paterova, Yang, An,
  Kalashnikov, and Krivitsky}}]{Paterova_2018}
\bibinfo{author}{\bibfnamefont{A.~V.} \bibnamefont{Paterova}},
  \bibinfo{author}{\bibfnamefont{H.}~\bibnamefont{Yang}},
  \bibinfo{author}{\bibfnamefont{C.}~\bibnamefont{An}},
  \bibinfo{author}{\bibfnamefont{D.~A.} \bibnamefont{Kalashnikov}},
  \bibnamefont{and} \bibinfo{author}{\bibfnamefont{L.~A.}
  \bibnamefont{Krivitsky}}, \bibinfo{journal}{Quantum Science and Technology}
  \textbf{\bibinfo{volume}{3}}, \bibinfo{pages}{025008} (\bibinfo{year}{2018}),
  \urlprefix\url{https://doi.org/10.1088%2F2058-9565%2Faab567}.

\bibitem[{\citenamefont{Lahiri et~al.}(2017{\natexlab{b}})\citenamefont{Lahiri,
  Hochrainer, Lapkiewicz, Lemos, and Zeilinger}}]{lahiri2017twin}
\bibinfo{author}{\bibfnamefont{M.}~\bibnamefont{Lahiri}},
  \bibinfo{author}{\bibfnamefont{A.}~\bibnamefont{Hochrainer}},
  \bibinfo{author}{\bibfnamefont{R.}~\bibnamefont{Lapkiewicz}},
  \bibinfo{author}{\bibfnamefont{G.~B.} \bibnamefont{Lemos}}, \bibnamefont{and}
  \bibinfo{author}{\bibfnamefont{A.}~\bibnamefont{Zeilinger}},
  \bibinfo{journal}{Phys. Rev. A} \textbf{\bibinfo{volume}{96}},
  \bibinfo{pages}{013822} (\bibinfo{year}{2017}{\natexlab{b}}).

\bibitem[{\citenamefont{Hochrainer et~al.}(2017)\citenamefont{Hochrainer,
  Lahiri, Lapkiewicz, Lemos, and Zeilinger}}]{hochrainer2017quantifying}
\bibinfo{author}{\bibfnamefont{A.}~\bibnamefont{Hochrainer}},
  \bibinfo{author}{\bibfnamefont{M.}~\bibnamefont{Lahiri}},
  \bibinfo{author}{\bibfnamefont{R.}~\bibnamefont{Lapkiewicz}},
  \bibinfo{author}{\bibfnamefont{G.~B.} \bibnamefont{Lemos}}, \bibnamefont{and}
  \bibinfo{author}{\bibfnamefont{A.}~\bibnamefont{Zeilinger}},
  \bibinfo{journal}{Proceedings of the National Academy of Sciences}
  \textbf{\bibinfo{volume}{114}}, \bibinfo{pages}{1508} (\bibinfo{year}{2017}).

\bibitem[{\citenamefont{Lahiri}(2018)}]{lahiri2018}
\bibinfo{author}{\bibfnamefont{M.}~\bibnamefont{Lahiri}},
  \bibinfo{journal}{Phys. Rev. A} \textbf{\bibinfo{volume}{98}},
  \bibinfo{pages}{033822} (\bibinfo{year}{2018}),
  \urlprefix\url{https://link.aps.org/doi/10.1103/PhysRevA.98.033822}.

\bibitem[{\citenamefont{Krenn et~al.}(2017)\citenamefont{Krenn, Hochrainer,
  Lahiri, and Zeilinger}}]{MarioPhysRevLett2017}
\bibinfo{author}{\bibfnamefont{M.}~\bibnamefont{Krenn}},
  \bibinfo{author}{\bibfnamefont{A.}~\bibnamefont{Hochrainer}},
  \bibinfo{author}{\bibfnamefont{M.}~\bibnamefont{Lahiri}}, \bibnamefont{and}
  \bibinfo{author}{\bibfnamefont{A.}~\bibnamefont{Zeilinger}},
  \bibinfo{journal}{Phys. Rev. Lett.} \textbf{\bibinfo{volume}{118}},
  \bibinfo{pages}{080401} (\bibinfo{year}{2017}),
  \urlprefix\url{https://link.aps.org/doi/10.1103/PhysRevLett.118.080401}.

\bibitem[{\citenamefont{Kysela et~al.}(2019)\citenamefont{Kysela, Erhard,
  Hochrainer, Krenn, and Zeilinger}}]{kysela2019experimental}
\bibinfo{author}{\bibfnamefont{J.}~\bibnamefont{Kysela}},
  \bibinfo{author}{\bibfnamefont{M.}~\bibnamefont{Erhard}},
  \bibinfo{author}{\bibfnamefont{A.}~\bibnamefont{Hochrainer}},
  \bibinfo{author}{\bibfnamefont{M.}~\bibnamefont{Krenn}}, \bibnamefont{and}
  \bibinfo{author}{\bibfnamefont{A.}~\bibnamefont{Zeilinger}},
  \bibinfo{journal}{arXiv preprint arXiv:1904.07851}  (\bibinfo{year}{2019}).

\bibitem[{\citenamefont{Bruschi et~al.}(2017)\citenamefont{Bruschi, Sab\'{\i}n,
  and Paraoanu}}]{Paraoanu_many-part-ent_PRA}
\bibinfo{author}{\bibfnamefont{D.~E.} \bibnamefont{Bruschi}},
  \bibinfo{author}{\bibfnamefont{C.}~\bibnamefont{Sab\'{\i}n}},
  \bibnamefont{and} \bibinfo{author}{\bibfnamefont{G.~S.}
  \bibnamefont{Paraoanu}}, \bibinfo{journal}{Phys. Rev. A}
  \textbf{\bibinfo{volume}{95}}, \bibinfo{pages}{062324}
  (\bibinfo{year}{2017}),
  \urlprefix\url{https://link.aps.org/doi/10.1103/PhysRevA.95.062324}.

\bibitem[{\citenamefont{Zhang et~al.}(2012)\citenamefont{Zhang, Bian, Chen, Ou,
  and Zhang}}]{zhang2012way}
\bibinfo{author}{\bibfnamefont{G.}~\bibnamefont{Zhang}},
  \bibinfo{author}{\bibfnamefont{C.}~\bibnamefont{Bian}},
  \bibinfo{author}{\bibfnamefont{L.}~\bibnamefont{Chen}},
  \bibinfo{author}{\bibfnamefont{Z.}~\bibnamefont{Ou}}, \bibnamefont{and}
  \bibinfo{author}{\bibfnamefont{W.}~\bibnamefont{Zhang}},
  \bibinfo{journal}{New Journal of Physics} \textbf{\bibinfo{volume}{14}},
  \bibinfo{pages}{063034} (\bibinfo{year}{2012}).

\end{thebibliography}

\end{document}